\documentclass[
  twocolumn,
  prb,
  showpacs,
  amsmath,
  amssymb,
  superscriptaddress
]{revtex4}

\usepackage{bm}
\usepackage{bbm}
\usepackage{graphicx}
\newcommand{\sign}{\mathop{\mathrm{sign}}}
\usepackage{hyperref}
\begin{document}

\title{Dynamics of waves in 1D electron systems:\\ Density oscillations
driven by population inversion}

\author{I.\ V.\ Protopopov}
\affiliation{
 Institut f\"ur Nanotechnologie, Karlsruhe Institute of Technology,
 76021 Karlsruhe, Germany
}
\affiliation{
 L.\ D.\ Landau Institute for Theoretical Physics RAS,
 119334 Moscow, Russia
}

\author{D.\ B.\ Gutman}
\affiliation{Department of Physics, Bar Ilan University, Ramat Gan 52900,
Israel }
\affiliation{
 Institut f\"ur Nanotechnologie, Karlsruhe Institute of Technology,
 76021 Karlsruhe, Germany
}

\author{P. Schmitteckert}

\affiliation{
 Institut f\"ur Nanotechnologie, Karlsruhe Institute of Technology,
 76021 Karlsruhe, Germany
}
\affiliation{
 DFG Center for Functional Nanostructures, Karlsruhe Institute of Technology,
 76128 Karlsruhe, Germany
}

\author{A.\ D.\ Mirlin}

\affiliation{
 Institut f\"ur Nanotechnologie, Karlsruhe Institute of Technology,
 76021 Karlsruhe, Germany
}
\affiliation{
 Inst. f\"ur Theorie der kondensierten Materie,
 Karlsruhe Institute of Technology, 76128 Karlsruhe, Germany
}
\affiliation{
 DFG Center for Functional Nanostructures, Karlsruhe Institute of Technology,
 76128 Karlsruhe, Germany
}

\affiliation{
 Petersburg Nuclear Physics Institute,
 188300 St.~Petersburg, Russia.
}


\begin{abstract}
We explore dynamics of a density pulse induced by a local quench in a
one-dimensional electron system. The spectral curvature leads to an
``overturn'' (population inversion) of the wave. We show that beyond this time
the density profile develops strong oscillations with a period much larger than
the Fermi wave length. The effect is studied first for the case of free fermions
by means of  direct quantum simulations and via semiclassical  analysis of the evolution of Wigner function. 
We demonstrate then that the period of oscillations is correctly reproduced by a hydrodynamic
theory with an appropriate dispersive term. Finally, we explore the effect of
different types of electron-electron interaction on the phenomenon. We show that sufficiently strong interaction
[$U(r)\gg 1/mr^2$ where $m$ is the fermionic mass and $r$  the relevant spatial scale] determines the dominant dispersive term in the hydrodynamic equations. Hydrodynamic theory reveals crucial dependence of the density evolution on the relative sign of the interaction and the density perturbation.

\end{abstract}


\maketitle

\section{Introduction}
\label{s1}

Transport properties of interacting one-dimensional (1D) systems keep
attracting a great deal of research interest. Experimental realizations of 1D
fermionic systems include, in particular, carbon nanotubes, semiconductor and
metallic nanowires, as well as quantum Hall (and other topological insulator)
edges. Further, 1D bosonic and fermionic systems can be engineered by using
cold atomic gases in optical traps of the corresponding geometry. A standard
and powerful theoretical approach to interacting 1D systems is the bosonization
\cite{stone,Delft,Gogolin,giamarchi,maslov-lectures}. When (i) the spectrum is
linearized, (ii) backscattering processes are neglected, and (iii) the physics
near equilibrium is explored, the bosonization reduces the original interacting
problem to a Gaussian field theory, thus reducing evaluation of physical
observables to a straightforward calculation of Gaussian integrals. 
When one (or several) of the above three conditions is not fulfilled, the
theoretical analysis becomes much more involved. In the present paper we will
focus on non-equilibrium physics of 1D fermionic systems in the regime where
the spectral curvature is of crucial importance. 

Properties of 1D interacting
systems with spectral nonlinearity have been addressed in a series of recent
theoretical works\cite{deshpande10,imambekov09,imambekov11,khodas07,karzig10}. 
Here we will consider the time evolution of a density pulse created
by a local quench in a 1D fermionic system (that will be assumed to be spinless
or spin-polarized for simplicity). We will assume that this pulse is
quasiclassical (i.e., has a characteristic spatial extension much larger than
the Fermi wave length) and sufficiently strong (i.e., contains a large number of
electrons). For not too long times the evolution seems to be fully harmless: the
pulse splits into left- and right-moving parts that separate and move away
from each other, approximately preserving theirs shape.  They key point is
that the shape would remain strictly unchanged only for linear dispersion of
excitations, while the non-linearity of dispersion leads to a 
deformation of the pulse. As a result, at a certain finite time the pulse tends
to ``overturn''. The problem to be addressed is what happens with the
density profile beyond this time. 

The above problem was formulated in Ref.~\onlinecite{bettelheim06} in the
context of Calogero model that was argued to describe the fractional quantum
Hall (FQH) edges (see also a recent paper Ref.~\onlinecite{wiegmann12}). Using
quantum hydrodynamics approach, the authors of
Refs.~\onlinecite{bettelheim06,wiegmann12} came to a conclusion that a density
pulse formed in a FQH edge will evolve in a sequence of well separated solitons
with a quantized charge (equal to $\nu$ for Laughlin  states).

In the present work we perform a systematic analysis of the pulse dynamics for
free fermions as well for those with different types of
interaction. We begin by considering a non-interacting case (Sec.~\ref{s2}). 
Quantum simulations show development of density oscillations at sufficiently large times.  
By analyzing evolution of the Wigner function, we show that once the semiclassical
phase-space distribution overturns (i.e. develops a population inversion
characterized by three ``Fermi momenta''), strong oscillations of density are
generated in the corresponding region of space. The characteristic scale of
these oscillations is much larger than the Fermi wave length $\lambda_F$. The
oscillations can be understood as Friedel oscillations between different
Fermi-momentum branches. 

In Sec.~\ref{s3} we switch to the bosonization language and discuss a connection
between free-fermion oscillations studied in Sec.~\ref{s2} and classical
hydrodynamics. In the latter class of problems \cite{whitham} oscillating
structures are known to develop when shock waves are regularized by dispersive
terms. We show that although dispersive terms arise already within Haldane
bosonization formalism of free fermions with curvature, the dominant terms
should come from summing the loop expansion. While we do not know how to take
into account these effects systematically, we approximate them by
including in the classical hydrodynamic equation a term corresponding to an
upper (for a positive pulse) branch of the particle-hole continuum. Solving the
corresponding equation (which is of Benjamin-Ono type), we show that it yields
oscillations with correct period (including its spatial variation) but with an
amplitude several times larger than the right one. This shows that the
above classical hydrodynamic equation does catch some important physics of
the developing ``dispersive shock'' of a free-fermion pulse but does not
represent a fully controllable approximation. 

Section \ref{s4} is devoted to an analysis of the interaction effects on
the pulse evolution. We consider first the case of a short-range interaction
and argue that the discovered oscillations remain largely preserved, up to two
modifications: (i) conventional Luttinger-liquid renormalization of the
Fermi velocity, and (ii) washing out of oscillations at long times due to
inelastic processes. 
We turn then to the case of a long-range interaction. We show that when the
interaction decays sufficiently slowly  (specifically, $U(r) \gg 1/mr^2$ at
large distances $r$, and $m$  is a electronic mass), the leading contribution to
the dispersion results from the interaction term, and the problem can
be treated quasiclassically (i.e. loops can be neglected), giving rise to a
classical hydrodynamic equation. The evolution of the pulse according to such
an equation depends crucially on the sign of the pulse and the sign of 
the interaction (or, more precisely, on the relative sign between them). When
the interaction is repulsive and the density pulse is downward,
oscillations develop similarly to the case of free fermions (or short-range
interaction). The period of oscillations gets however parametrically larger. On
the other hand, for an upward pulse (and still assuming a repulsive long-range
interaction), the pulse splits in a sequence of ``solitons'' (whose charge is
in general not quantized, except for the case of $1/r^2$ interaction).

Section \ref{s5} contains a summary of our results and a discussion of
prospective research directions.

\section{Free fermions}
\label{s2}
\begin{figure}
 \includegraphics[width=210pt]{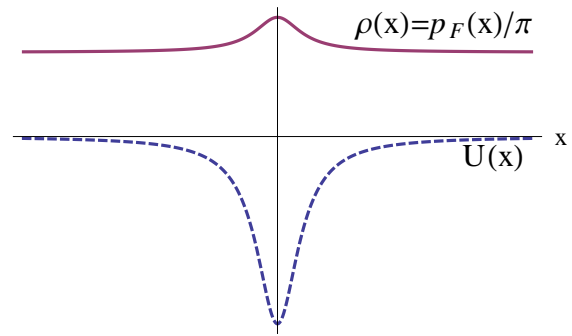}
\caption{\small Setup. Density disturbance in the Fermi sea is
created by the application of a potential $U(x)$ which is then switched off at
$t=0$.  }
\label{Fig:Ur}
\end{figure}

\begin{figure*}
\includegraphics[width=160pt]{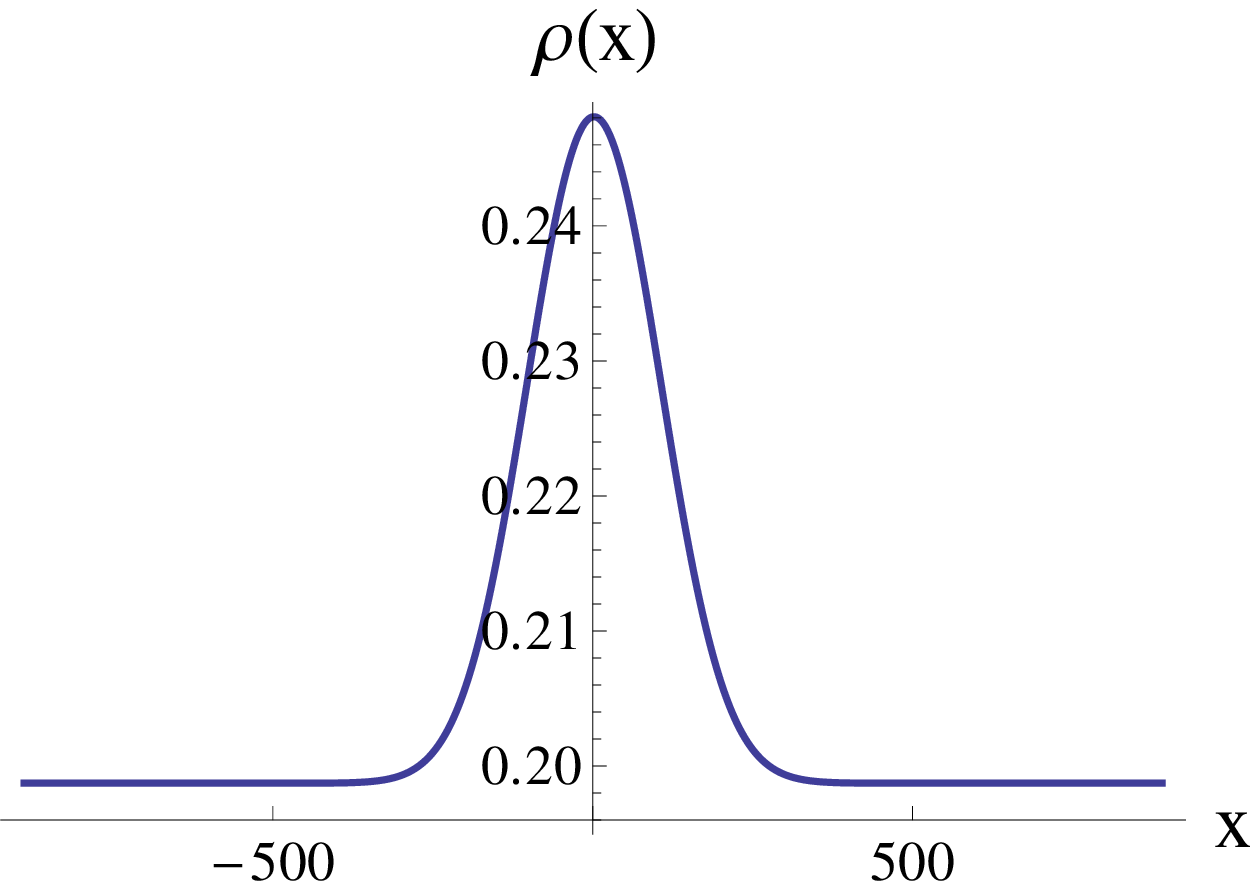}
\includegraphics[width=160pt]{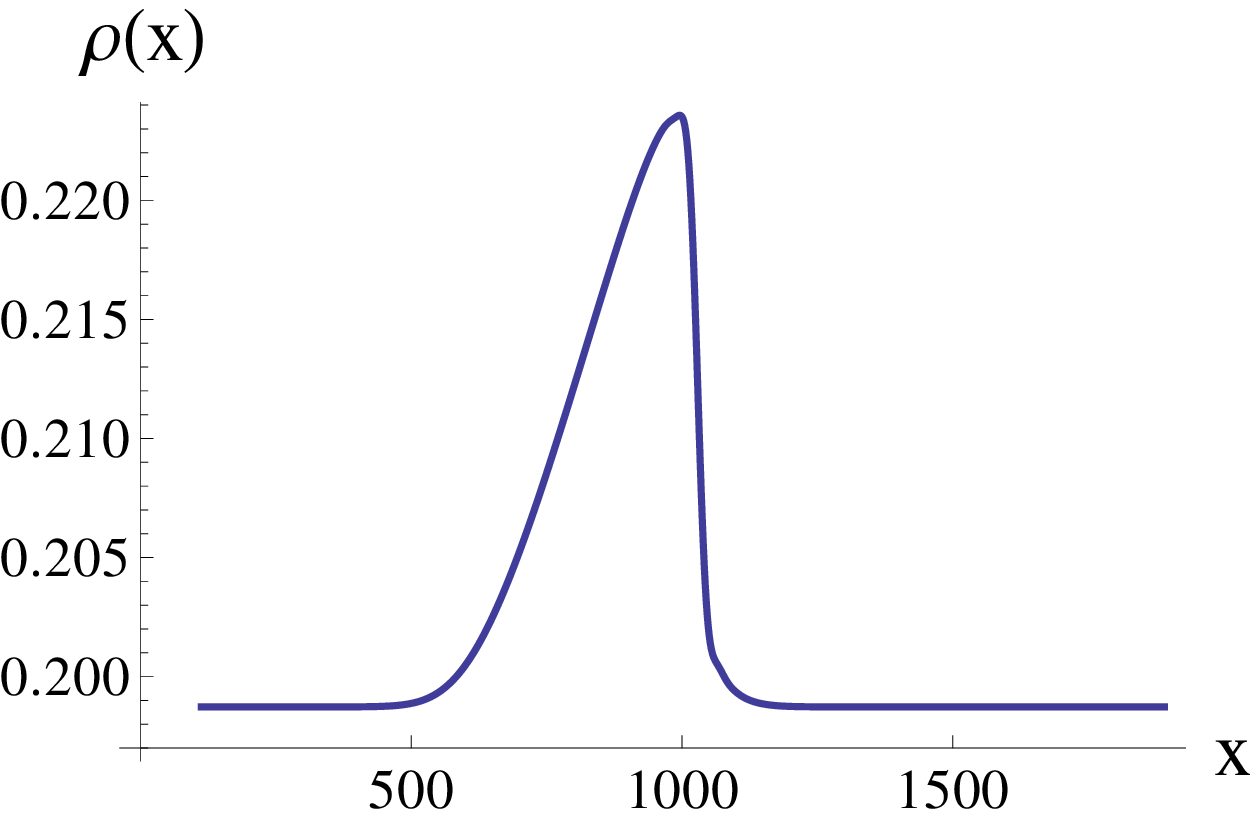}
\includegraphics[width=160pt]{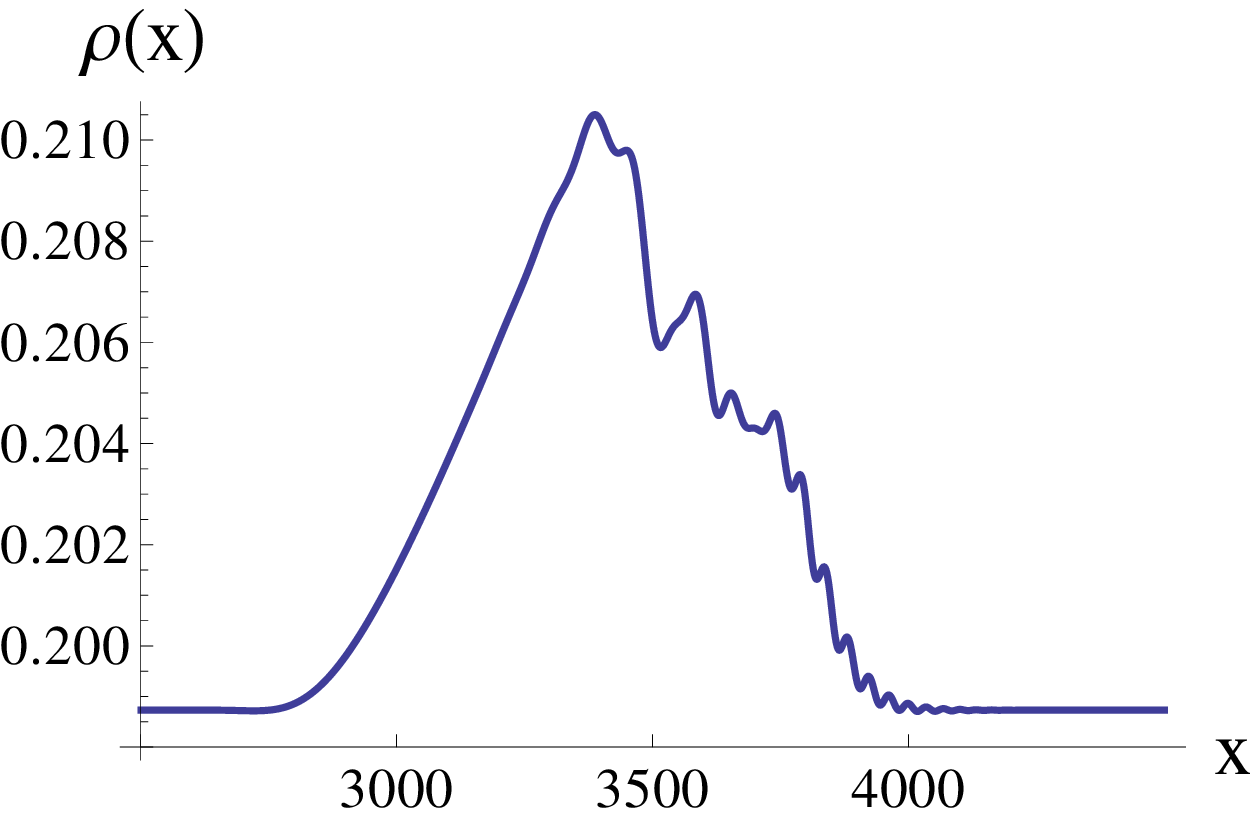}
\caption{\small  
Evolution of the fermionic density (measured in fermions per site)  as obtained from direct numerical simulations of free fermions on a lattice. The unit of length is set by the lattice constant.  
{\it Left:} The initial density pulse has Gaussian shape with the amplitude $\Delta\rho \approx 0.05$ and
the dispersion $\Delta x \approx 102.6$, comprising 12.8 electrons.
{\it Middle:} 
Density profile at $t$ slightly smaller than the ``shock'' time $t_c$. Only right-moving part is shown.
The front edge of the pulse has become steep but the overturn has not yet occurred. 
{\it Right:} Fermionic density after the ``shock'', $t \approx 5t_c$. Density ripples develop at front edge.    
 }
\label{Fig:Peter}
\end{figure*}

In this section and in Sec.~\ref{s3} we study the evolution of a
``quasiclassical'' density
disturbance 
of the Fermi sea of free fermions.  We begin  (Sec.~\ref{s2.1}) by formulating
the problem and performing its numerical modelling which shows emergence
of density oscillations after the time corresponding to
overturning of the initial packet. In Sec.~\ref{s2.2} we
solve this problem analytically by using
the straightforward (``fermionic'') approach. Specifically, we demonstrate
that, once the dispersion induces a population inversion within the pulse, 
phase-space oscillation of the Wigner function give rise to density
oscillations. Analyzing the resulting density oscillations, we find a perfect
agreement with the results of numerical simulations of Sec.~\ref{s2.1}. 
Finally, in Sec.~\ref{s3} we make a link to ``dispersive shocks'' in
the classical hydrodynamics. We show that when the dispersive term corresponding
to the appropriate branch of the particle-hole spectrum is incorporated into
classical hydrodynamic equations, the latter reproduce correctly the period of
emerging oscillations (but considerably overestimates their amplitude).

\subsection{Formulation of the problem and numerical simulations}
\label{s2.1}

\begin{figure}
\includegraphics[width=223pt]{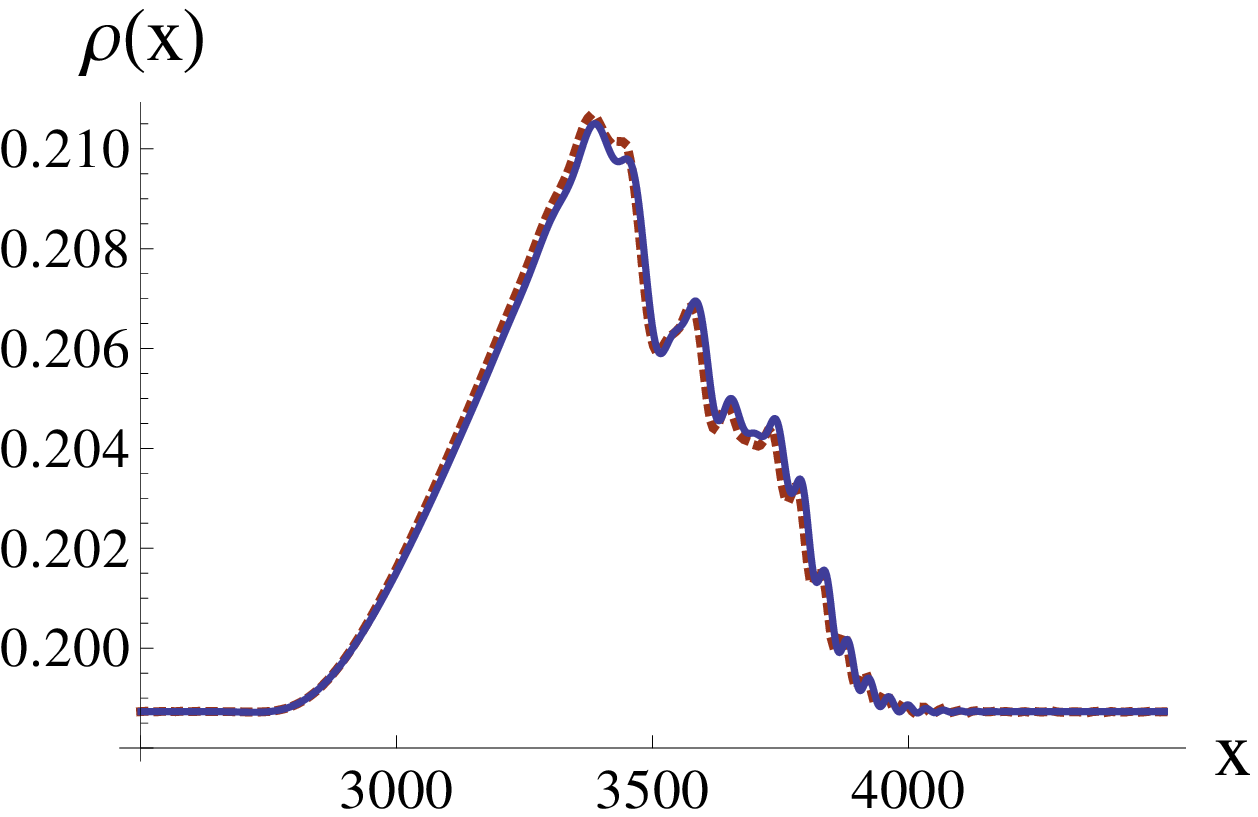}
\includegraphics[width=223pt]{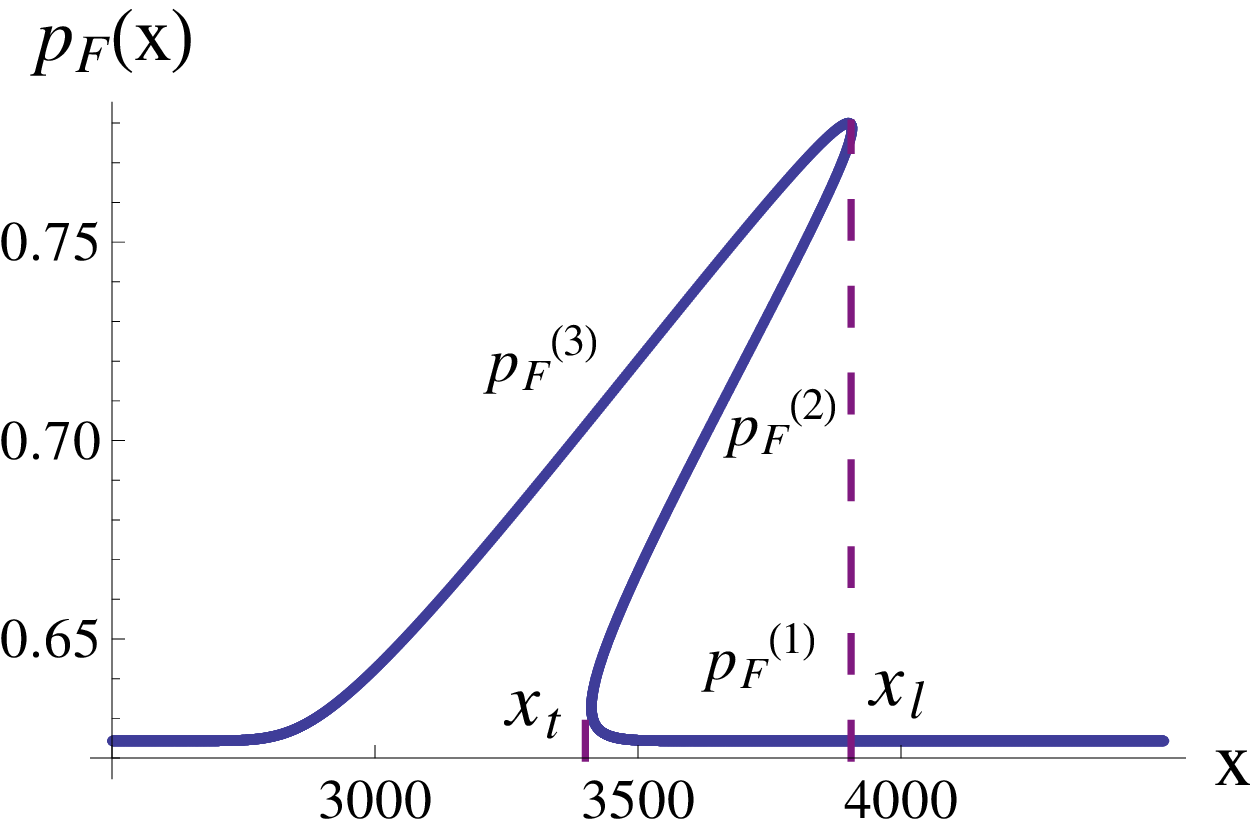}
\includegraphics[width=223pt]{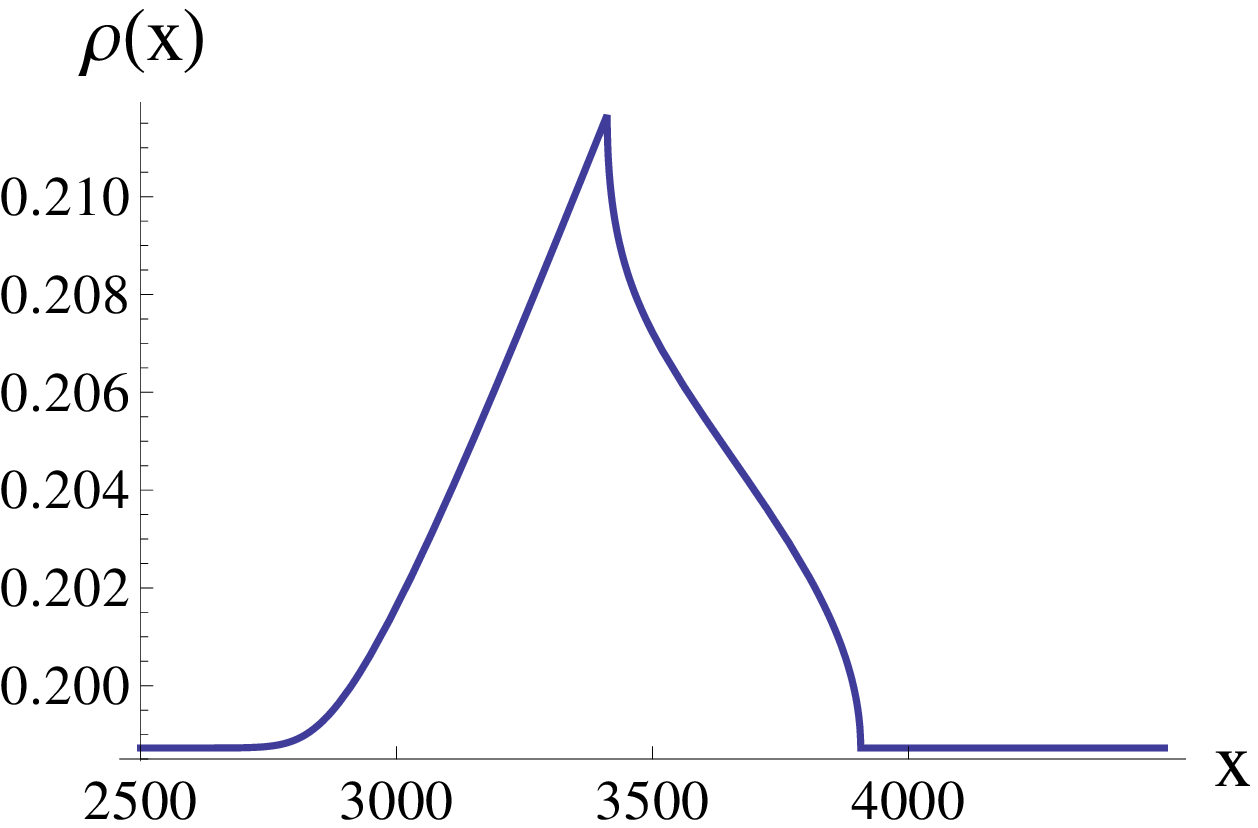}
\caption{\small
{\it Top:}
Snapshot of   fermionic density $t\approx 5t_c$ as obtained from direct numerical simulations of free fermions on a lattice. The initial density pulse was Gaussian (see main text and Fig.~\ref{Fig:Peter}). 
The dots show the semiclassical result derived in Sec. \ref{s2.2}.
 {\it Middle:}
 Quasiclassical phase-space distribution as 
obtained from the Euler equation at $t\approx 5t_c$.
   In between trailing and leading edges $x_t$ and $x_l$ the Fermi surface 
is multivalued with branches $p_F^{(i)}\,, \; i=1, 2, 3$. 
 {\it Bottom:} Naive quasiclassical
approximation to the density profile as obtained by momentum integration of the
phase-space distribution shown in the top panel. }
\label{Fig:pFBolzman}
\end{figure}

The problem that we address is formulated in a rather simple way. 
We assume that a non-uniform fermionic density was
created by application of a smooth (on the scale of $\lambda_F$)
and relatively weak (compared to the Fermi energy $\epsilon_F$) external 
potential $U(x)$ to the unperturbed Fermi sea (Fig.~\ref{Fig:Ur}).  The system
at $t<0$ is in its ground state characterized by the fermionic density
\begin{equation}
 \rho_0(x)\equiv \frac{p_F(x)}{\pi}=\frac{1}{\pi}\sqrt{p_\infty^2-2mU(x)}\,.
\label{Eq:rho_0}
\end{equation}
Here, $p_\infty$ is the Fermi momentum at infinity. 
Note that all the corrections to the semi-classical result (\ref{Eq:rho_0}) are exponentially small as long as 
$U(x)$ is smooth on the scale of $\lambda_F$. For
transparency of discussion we assume that the density pulse has a shape
of a single hump (as shown schematically in Fig.~\ref{Fig:Ur}),
i.e. that $U(x)<0$  and has a single minimum at $x=0$.

At $t=0$ the potential is suddenly switched off, which results in the
appearance of a non-equilibrium state and subsequent 
propagation of the density perturbation created by $U(x)$.
Our goal will be to explore this density evolution at sufficiently long times.
We will assume that the number of particles within the initial pulse is large,
$\Delta x \: \Delta\rho \gg 1$, where $\Delta x$ and $\Delta \rho$ are the
characteristic extension of the pulse and its amplitude, respectively. The
interesting physics will emerge at times $t > t_c \sim m \Delta x/\Delta\rho$,
when the semiclassical phase-space distribution overturns.

Since the fermions are free and their state was prepared in the coherent manner described above,  
the full information on the quantum state of the system is encoded in the 
Wigner function 
\begin{eqnarray}
 f(X, p; t)=\int dy e^{-ipy}\left\langle\psi^+\left(X_-; t\right)
\psi\left(X_+; t\right)\right\rangle\,,
\label{Eq:f_def}
\\
X_\pm=X\pm \frac{y}{2}\qquad\qquad\qquad
\nonumber
\end{eqnarray}
satisfying at $t>0$ the (exact!) Boltzmann equation
\begin{equation}
 \partial_t f(X, p; t)+p\partial_X f(X, p; t)=0\,.
\label{Eq:Boltzmann}
\end{equation}
In this equation  we have set the particle mass to unity. 
Let us note that the mass $m$ will enter our results only through the overall time scale. Thus, dependence  on $m$  can be eliminated completely by measuring $t$ in units of~$t_c$.  

Equation (\ref{Eq:Boltzmann}) is trivially solved, yielding
\begin{equation}
 f(X, p; t)=f\left(X-p t, p; t=0\right)\equiv f_0\left(X-p t, p\right)\,.
\label{Eq:f_t}
\end{equation}
Therefore, once we know the Wigner function of the initial ($t=0$) state, the
Wigner function of the evolved ($t>0$) state can be immediately obtained. 
 
Let us examine now the Wigner function of the initial state.
Semiclassically one would expect that $f_0(X, p)$ takes value unity for the
occupied electronic states that are below the position-dependent Fermi
momentum $p_F(X)$ and is zero for empty ones (above $p_F(X)$):
\begin{equation}
 f_0(X, p)=\Theta\left(p_F^2(X)-p^2\right)\,.
\label{Eq:f_0_trivial}
\end{equation}
In this approximation, the time-dependent state of the fermions after the quench
(i.e. at $t>0$) is fully characterized by the Fermi surface
$p_F(x, t)$ separating occupied and unoccupied single-particle states in the
phase space and satisfying the Euler equation
(we concentrate on the Fermi surface for the right-moving particles with $p_F(x)>0$)
\begin{equation}
 \partial_t p_F(x, t)+p_F(x, t)\partial_x p_F(x, t)=0\,.
\label{Eq:Euler}
\end{equation}

While  capturing correctly the physics at small times, the Euler equation
(\ref{Eq:Euler}) suffers from the shock-wave phenomenon. Specifically, for
arbitrarily smooth initial conditions, the curvature of
the electronic dispersion relation $\epsilon(p)=p^2/2$, makes
the Fermi surface  $p_F(x)$ multivalued at large enough times ($t>t_c$) and
leads to the appearance 
of infinite spatial gradients of fermionic density (Fig.~\ref{Fig:pFBolzman}, middle and bottom panels). 
This suggests that the simple semiclassical description
(\ref{Eq:Euler}), (\ref{Eq:f_0_trivial}) may become insufficient beyond the
time $t_c$ when the shock occurs, raising the question of what happens
with the density profile at~$t>t_c$. 

We have performed direct quantum simulations of this problem by using a
tight-binding  free-fermion model. Figure~\ref{Fig:Peter} demonstrates the
 density evolution from initial state at $t=0$ (left panel) to state at   certain time after the shock 
(approximately five times larger than $t_c$).
At $t>0$ only the right-moving part of the density pulse is
shown. A full movie of density evolution is available online~\cite{footnote-movie}.  
  The initial density perturbation  was Gaussian 
\begin{equation}
 \rho_0(x)=\rho_\infty+\frac{N}{\sqrt{2\pi \sigma^2}}e^{-x^2/2\sigma^2}
\label{pulse_Gaussian}
\end{equation}
with dispersion $\sigma$ of about $100$ lattice sites and contained  $N\approx 12.8$
electrons. The density of the underlying Fermi sea is $0.2$ fermion per site, so
that the cosine-shaped dispersion relation of the tight-binding model can be
well approximated by a parabola.

For convenience of the reader the snapshot of the density at $t\approx 5t_c$ is also shown in the top panel of Fig.~\ref{Fig:pFBolzman}.
Comparing the exact quantum result (Fig.\ref{Fig:pFBolzman}, top)  to the naive semiclassical result dictated by the Euler
equation (Fig.~\ref{Fig:pFBolzman}, bottom), we see that the shock gets regularized via
the onset of pronounced density oscillations at the front edge of the pulse. 
It is important to emphasize that the period
of those oscillations is controlled by the amplitude $\Delta\rho$ of the density
perturbation and is thus much larger than $\lambda_F$.  (We will
perform a detailed quantitative analysis of the oscillation period below in
Sec.~\ref{s2.2}.) From this point of view, the developing oscillations may be
considered as quasiclassical: their characteristic scale is much larger than
$\lambda_F$. Thus, a smooth initial density stays smooth at scale $\lambda_F$ 
also at times after the ``shock''.

We thus face an apparent contradiction: the density profile remains
``quasiclassical'' (smooth on the scale of $\lambda_F$) after the shock but
develops strong oscillations that are not caught by the
quasiclassical approximation based on Eqs.~(\ref{Eq:Euler}),
(\ref{Eq:f_0_trivial}). The resolution of this ``paradox'' is related to the
fact that Eq.~(\ref{Eq:f_0_trivial}) is not the fully correct semiclassical
(in the above sense) approximation for the Wigner function  of fermions in a
smooth potential well. As was pointed out in
Ref.~\onlinecite{bettelheim11}, instead 
of having abrupt drop from $1$ to $0$ at Fermi momentum, $f_0(X, p)$ as a
function of $p$ develops oscillations near $p_F(X)$. Those oscillations can be
considered as a semiclassical effect in the sense that their form knows nothing
about $\lambda_F$ and is controlled solely by the derivatives  of $p_F(X)$. In
Sec.~\ref{s2.2} we give a detailed account of the oscillations in 
the Wigner function and of their implications for the density
evolution.   

\subsection{Wigner function of fermions in a potential well and density
oscillations}
\label{s2.2}

The Wigner function $f_0(p, X)$ of the initial state satisfies the equation
\begin{equation}
\partial_X\partial_y
f_0(X,y)-\left[U(X_+)-U(X_-)\right]
f_0(X, y)=0\,,
\label{Eq:Eq_f_0_general}
\end{equation}
where the  coordinate $y$ is conjugate (in the sense of Fourier
transformation) to the momentum $p$, cf. Eq.~(\ref{Eq:f_def}).
As we are interested in the behavior of $f_0(p, X)$ close to one of the Fermi edges $p=\pm p_\infty$ we can replace 
$\partial_y$ by $ip_\infty$ (we concentrate here on the right Fermi edge).
This corresponds to taking the limit $p_\infty \to \infty$ while keeping
the profile $p_F(x)-p_\infty$ fixed.
Solving the resulting equation
\begin{equation}
 \partial_X f_0(X, y) -i 
\left[p_{F}(X_+)-p_{F}(X_-)\right]f_0(X, y)=0\,,
\end{equation}
with the condition at infinity 
\begin{equation}
 f_0(X=-\infty, y)=\int \frac{dp}{2\pi}e^{ipy}\Theta\left(p_\infty-p\right)
\end{equation}
 and transforming the result to momentum space, we find in agreement with
Ref.~\onlinecite{bettelheim11}
\begin{eqnarray}
 f_0(X, p)=\int\frac{dy}{2\pi i (y-i0)}e^{-iS[y; X, p]}\,,
\label{Eq:f_0_Correct}
\\
S[y; X, p]=py-\int_{X-\frac{y}{2}}^{X+\frac{y}{2}}dX' p_F(X')\,.
\label{Eq:S[y; X, p]}
\end{eqnarray}
Note that our approach is slightly different from that of
Ref.~\onlinecite{bettelheim11}: we consider an equilibrium state in a potential
$U(x)$, while the authors of Ref.~\onlinecite{bettelheim11} construct a coherent
state by acting on the homogeneous Fermi vacuum  with an exponential of a
bilinear in fermionic operators. 

The general structure of the Wigner function can be inferred from
Eqs.~(\ref{Eq:f_0_Correct}), (\ref{Eq:S[y; X, p]})
by performing the integration with making use of the saddle-point method.
The saddle-point equation for the action (\ref{Eq:S[y; X, p]})
\begin{equation}
 p=\frac{p_F(X+y/2)+p_F(X-y/2)}{2}
\end{equation}
has no real-valued solutions for $p<p_\infty$ or 
\begin{equation}
p>p_{\rm max}(x)\equiv \max_{y}\frac{p_F(X+y/2)+p_F(X-y/2)}{2}\,.
\end{equation}
In these parts of the phase space the integral (\ref{Eq:f_0_Correct}) is
controlled by the singularity at $y=0$, and 
$f_0(X, p)$ can be well approximated by Heaviside $\Theta$-function. 

On the contrary, for $p_\infty<p<p_{\rm max}(X)$ there exist (at least) two
solutions $y=\pm y^*(X, p)$ to the saddle point equation providing an
oscillatory contribution to the Wigner function
\begin{equation}
 \delta f_0(X, p) \propto \Re\left[A e^{-iS[X, p]}\right]\,.
\end{equation}
 Here $A$ is the coefficient controlled by the fluctuations around the
saddle points.
The phase of oscillations is given by  $S[X, p]=S[y^*(X, p); X, p]$. It defines
the period of 
oscillations with the momentum via $S[X, p+\Delta p]-S[X, p]=2\pi$. Counting the
powers of imaginary unit in the saddle point   
integration one easily finds that the maxima of the Wigner function appear at
$S[X, p]=-3\pi/4 \mod 2\pi$. 
 
In a sufficiently close vicinity of the local Fermi surface, $|p-p_F(x)| \ll
p_F(0) - p_\infty$ one can locally approximate $p_F(x)$ by a
parabola and  express the Wigner function $f_0(x,p)$ in terms
of the Airy function \cite{bettelheim11}. This approximation is however
insufficient for our purposes, as the density oscillations will originate from
Wigner function oscillations at all scales $|p-p_F(x)| \sim
p_F(0) - p_\infty$.  
 
\begin{figure}
 \includegraphics[width=245pt]{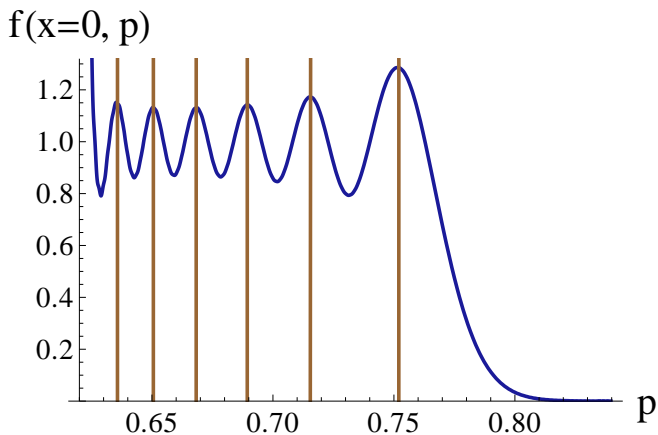}
\caption{\small Initial-state Wigner function $f_0(X=0, p)$ as obtained via
numerical integration of (\ref{Eq:f_0_Correct}). The same Gaussian
density profile as used in the quantum simulations presented in
Sec.~\ref{s2.1} and in Fig.~\ref{Fig:Peter} was assumed. The observed six
oscillations correspond to 6.4 right-moving particles in the pulse. Vertical
lines in Fig.~\ref{Fig:WignerFunctionAtZero} mark the momenta satisfying the
condition for oscillation maxima, $S[X, p]=-3\pi/4 \mod 2\pi$. }
\label{Fig:WignerFunctionAtZero}
\end{figure}

The oscillatory behavior of $f_0(X, p)$ is illustrated in
Fig.~\ref{Fig:WignerFunctionAtZero} where we plot $f(X=0, p)$  
(calculated via numerical integration of (\ref{Eq:f_0_Correct}))
as a function of momentum. To generate the plot we have assumed the Gaussian
density used in the quantum simulations presented in Sec.~\ref{s2.1} and in
Fig.~\ref{Fig:Peter}. A straightforward analysis of the saddle-point equation
shows that, upon variation of momentum from $p=p_{\rm max}(0)=p_F(0)$ to
$p_\infty$,
$S[X=0, p]$ varies monotonically from $0$ to $-2\pi N$, where 
\begin{equation}
N=\frac{1}{2\pi}\int dx  (p_F(x)-p_\infty)
\end{equation}
is (generally non-integer) number of particles in the pulse. Accordingly,
the Wigner function of Fig. \ref{Fig:WignerFunctionAtZero} shows $6$
oscillations corresponding to approximately $6$ right-moving particles. Vertical
lines in Fig.~\ref{Fig:WignerFunctionAtZero} mark the momenta satisfying the
condition $S[X, p]=-3\pi/4 \mod 2\pi$.  

\begin{figure}
 \includegraphics[width=245pt]{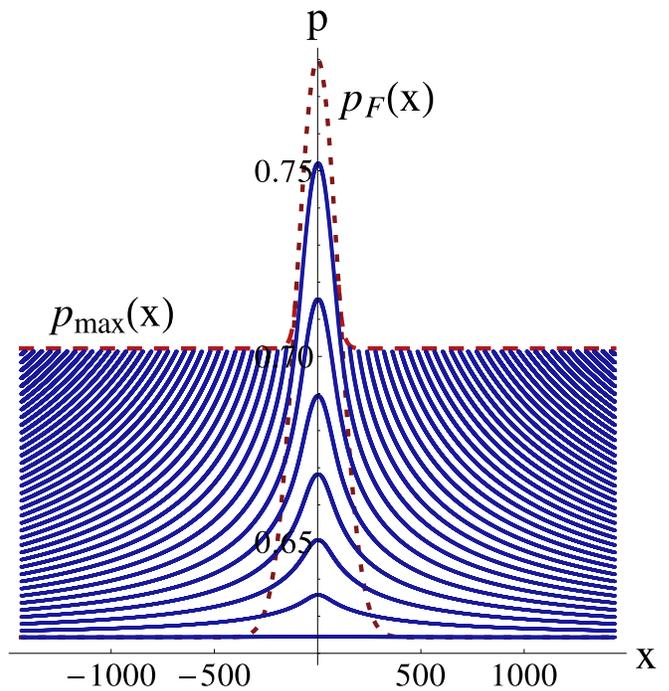}
\caption{\small Phase-space oscillations
of the Wigner function $f_0(x,p)$ for the initial state (the same Gaussian
density hump as in the previous figures; the corresponding $p_F(x)$ is shown by
dotted line). Blue solid lines are contours of constant action (which
determines the phase of the Wigner function), $S[X,p]=-3\pi/4 \mod 2\pi$. Dashed line 
represents the border $p_{\rm max}(x)$ of  the region of developed oscillations.
 }
\label{Fig:WiegnerFunctionInitial}
\end{figure}

The overall behavior of $f_0(X, p)$ in the phase space can be conveniently
represented by lines of constant  
action $S[X, p]$. For the case of Gaussian density the corresponding pictures is
shown on Fig.~\ref{Fig:WiegnerFunctionInitial}. The green dashed line here
represents the border $p_{\rm max}(x)$ of  the region of developed oscillations.
The solid blue lines are the lines of constant action $S[X,p]=-3\pi/4 \mod
2\pi$. Finally, the dotted line shows the $x$-dependent Fermi level. The
``topology'' of the plot can be understood on general grounds and does not
depend on the specific density $\rho_0(x)$. At large $X$ the action is a steep
function of momentum $p$ and, when taken modulo $2\pi$, 
acquires any given value many times. Among the lines of constant action coming
from $X=+\infty$, exactly $\left[N\right]$ (this denotes the integer part of
$N$) lines cross the $p$-axes and flow to $X=-\infty$, while other lines end up
on the border $p=p_{\rm max}(x)$ where the solution $y^*(X, p)$ to the saddle
point equation becomes complex.  

Let us now discuss the implications of the above results for the
fermionic density which is equal to the integral over
momentum of the Wigner function. In the initial state the density is insensitive
to the oscillations of the Wigner function. 
Indeed, one can observe that the contour of momentum integration
(vertical line on Fig. \ref{Fig:WiegnerFunctionInitial}) crosses many contours
of constant $S[X, p]$ but does not touch any of them, so that there is no
stationary-point contribution to the integral. 
In fact, evaluating the $p$ integral of Eq.~(\ref{Eq:f_0_Correct}), we find that
in the considered (large $p_F$) approximation the equality
$\rho_0(x)=p_F(x)/\pi$ is exact. 

\begin{figure}
 \includegraphics[width=245pt]{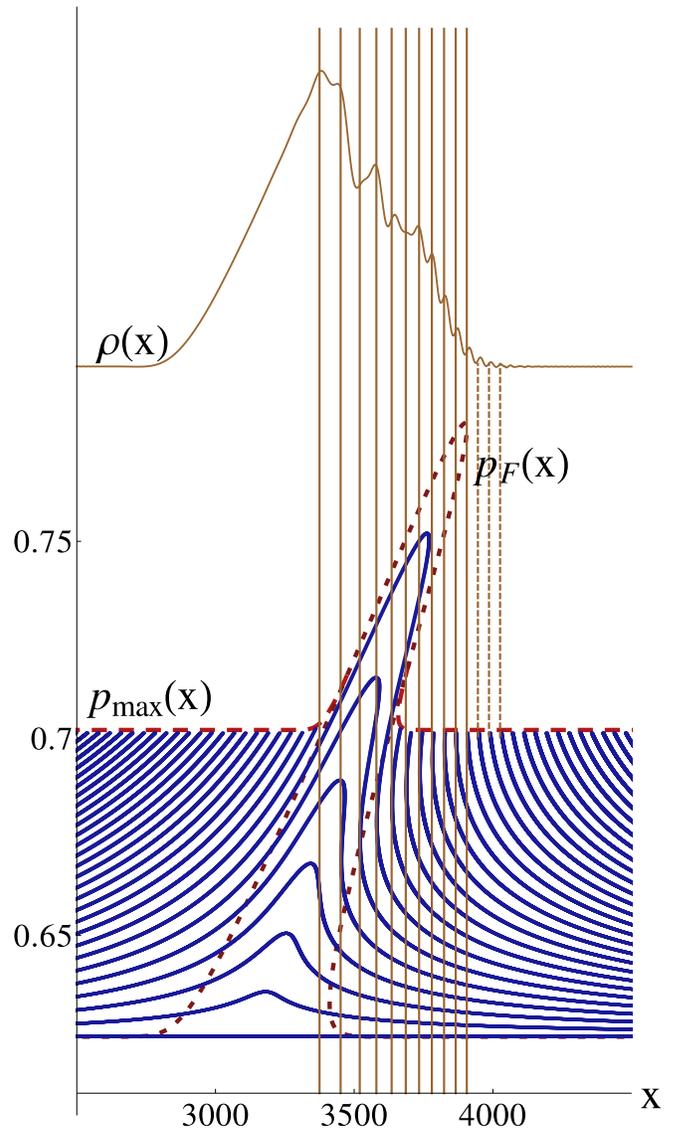}
\caption{\small {\it Top:} Density profile after the shock as obtained by
numerical numerical integration of the Wigner function
(\ref{Eq:f_0_Correct}). The result is almost indistinguishable from that
of first-principle quantum simulations shown in Fig.~\ref{Fig:pFBolzman}.
{\it Bottom:} The same as in Fig.~\ref{Fig:WiegnerFunctionInitial} but after
evolution beyond the shock. All parameters, including the evolution time, are
the same as in the upper plot and in Fig.~\ref{Fig:pFBolzman}. The vertical lines
touching the constant-action contours mark stationary-point
contributions to the momentum integral of the Wigner function and thus determine
maxima of the density oscillations in the upper plot.}
\label{Fig:WiegnerFunctionFinal}
\end{figure}

The  evolution of each contour line of the action is governed by the Euler
equation (\ref{Eq:Euler}). The behavior 
of the Wigner function after the shock,  $t>t_c$, is illustrated by
Fig.~\ref{Fig:WiegnerFunctionFinal}. Now the vertical lines 
do touch the contours of constant action. A touching point becomes the saddle
point for the integration over momentum and 
the oscillations in $f_0(X, p)$ start to contribute to the density. This implies
a maximum in the density 
each time the integration line touches the contour line corresponding to $S[X, p]=-3\pi/4 \mod 2\pi$.

The brown curve in the upper part of Fig.~\ref{Fig:WiegnerFunctionFinal} shows
the density obtained via numerical integration of the Wigner function
(\ref{Eq:f_0_Correct}). This is almost indistinguishable from the result of
first-principle quantum simulations (top panel of Fig.~\ref{Fig:pFBolzman}). We observe that the
positions of the maxima of the density are in accord with the above argument
based on the saddle-point approximation.

One can now determine the period of the density modulations. Let us focus
on the region closer to the front edge of the pulse where the oscillations are
clearly governed by a single harmonics, see Fig.~\ref{Fig:pFBolzman}, and are
perfectly described by the saddle-point argument as shown in
Fig.~\ref{Fig:WiegnerFunctionFinal}. As this figure further illustrates, the
integration contour touches the contour lines 
near $p=p_{\rm max}(X)$ and the period of the density oscillations is set by
$S[X, p=p_{\rm max}(X)]$. In this regime we have 
\begin{eqnarray}
&& y^*(X, p_{\rm max}(X)) \simeq 2X\,, \nonumber \\
&& p_{\rm max}(X) \simeq \frac{p_F(0)+p_\infty}{2} \,.
\end{eqnarray}
Thus, $ S[X, p_{\rm max}(X)]\simeq (p_F(0)-p_\infty)X$ 
and we immediately infer the period 
\begin{equation}
 \delta X \simeq \frac{2\pi}{p_F(0)-p_\infty}\,.
 \label{Eq:period}
\end{equation}

It is not difficult to generalize this argument to the region closer to the top
of the pulse. Assuming for simplicity that the time that has passed after shock
is of the order of the shock time $t_c \sim m \Delta x/\Delta \rho$,  we find
that the characteristic spatial scale for the first few
oscillations (just to the right of the maximum of the pulse) is larger
than (\ref{Eq:period}) by a factor $\sim N^{1/3}$ (for the parameters used in
our plots this is approximately two). Some complication comes from the fact
that in this region a superposition of oscillations originating from different
regions in phase space takes place. Indeed, as is clearly seen in the lower
panel of Fig.~\ref{Fig:WiegnerFunctionFinal}, constant-action contours that
do not terminate at $p_{\rm max}(x)$ may have two points with infinite slope,
and each of them will give a stationary-phase contribution when the momentum
integration is performed. This superposition explains a somewhat irregular
oscillation pattern in the corresponding spatial region.  

The obtained oscillations can be interpreted as Friedel-type
oscillations between different branches of the Fermi momentum (that becomes
multivalued after the ``shock''). In particular, in the
front region, the upper two branches are close to the maximum value, $p_0(0)$,
while the lower branch is essentially equal to $p_\infty$, which yields exactly
Eq.~(\ref{Eq:period}).

\section{Hydrodynamics of free fermions}
\label{s3}

The analysis of the previous section provides a detailed description of the
evolution of coherent perturbation in the density of free fermions. The analysis
is complete and, as was also confirmed by numerical simulations,
essentially exact under our basic assumptions. It is appealing however to try to
formulate a hydrodynamic description of the problem which, in contrast to the
fermionic approach utilizing the notion of Wigner function,  would involve as
fundamental  objects only the density and the velocity of the electronic fluid.
Indeed, the hydrodynamics (bosonization) constitutes a convenient and powerful
framework for the discussion of interaction effects (to be considered in
Sec.~\ref{s4} below) which are otherwise hard to access. 

Usually, hydrodynamics rests on the assumption of local equilibrium forced by the particle collisions which wash out any 
features in the particle distribution function. In the present problem no such
equilibrium exists. Moreover, we saw above 
that the oscillating behavior of $f(p)$ is crucial for the density ripples observed in the shock region. Hence, 
one can expect that the evolution of the quantum-coherent many-particle state
can not be controllably described by classical equations of hydrodynamic type.
Despite this fact, one can ask if it is possible to design phenomenological
hydrodynamic equations 
which would capture qualitative features of the true density evolution. In this section we show that this is indeed possible and such phenomenological equations provide important insight into the physics of non-equilibrium many-particle system.

In our search for hydrodynamics it is convenient to start from the Euler equation~(\ref{Eq:Euler}) corresponding to 
the neglect of all oscillatory features in the Wigner function $f_0(X, p)$. Combined with analogous equation for the 
Fermi surface of left electrons at $t<t_{c}$ it can be rephrased in terms   of the mean density   and velocity
of the fluid as
\begin{eqnarray}
 \partial_t \rho+\partial_x(\rho v)=0\,,\qquad
  \partial_tv+\partial_x\left(\frac{v^2}{2}+\frac{\pi^2\rho^2}{2}\right)=0\,,\label{Eq:Euler1}\\
\rho(x)=\int\frac{dp}{2\pi}f(x, p)\,,\quad v(x)=\frac{1}{\rho(x)}\int\frac{dp}{2\pi}pf(x, p)\,.
\end{eqnarray}
Of course, equations (\ref{Eq:Euler1}) suffer from the same shock phenomenon as the original equation (\ref{Eq:Euler}).
Phenomenologically, we would like to add some terms to  Eq. (\ref{Eq:Euler1}) regularizing the shock instability.
We know that regularization goes trough the onset of density ripples in the shock region. 
This phenomenon is well known in hydrodynamics and usually referred to as
``dispersive regularization''\cite{hoefer06}.   It takes place when the shock
caused by non-linearity gets regularized by higher order derivatives consistent
with
the time reversal invariance of the equations. The classical examples are the
Korteweg-de-Vries (KdV) and Gross-Pitaevskii equations. 
It is the requirement of time reversal that  makes the ``dispersive
regularization'' very different from ``dissipative regularization'' achieved by
the introduction into the system of some type of viscosity. Examples of the
latter type are Navier-Stokes and Burgers equations. 

Let us now point out that Eqs.~(\ref{Eq:Euler1}) appear in the theory of
Fermi gas in yet another context and with slightly different meaning.
Specifically, the standard bosonization procedure applied to the fermions with
quadratic spectrum leads to Hamiltonian \cite{schick68, Sakita,Jevicki_Sakita}
\begin{equation}
\hat{H}=\int dx\left(\frac{\widehat{\rho}\widehat{v}^2}{2}+\frac{\pi^2\widehat{\rho}^3}{6}\right)\,.
\label{Eq:QuntumHamiltonian}
\end{equation}
Here, $\widehat{\rho}$ and $\widehat{v}$ are {\it operators} with the commutation relations
\begin{equation}
 \left[\widehat{\rho}(x), \widehat{v}(y)\right]=-i\delta'(x-y)\,.
\end{equation}
The Hamiltonian and the commutation relations imply the operator equations of
motion usually referred to as ``quantum Euler equations''
\begin{equation}
 \partial_t \widehat{\rho}+\partial_x(\widehat{\rho}\widehat{ v})=0\,,\qquad
\partial_t\widehat{v}+\partial_x\left(\frac{\widehat{v}^2}{2}+\frac{
\pi^2\widehat{\rho}^2}{2}\right)=0\,.
\label{Eq:Euler_Quantum}
\end{equation}

One can now see that there exist two sources of corrections to hydrodynamic
equations (\ref{Eq:Euler1}). First, there can be corrections
to the quantum Hamiltonian (\ref{Eq:QuntumHamiltonian}) missed by
the bosonization in its simplified form. If present, they would yield a direct
contribution to the quantum Euler equations (\ref{Eq:Euler_Quantum}). Second,
passing from quantum equations 
(\ref{Eq:Euler_Quantum}) to identically-looking classical equations
(\ref{Eq:Euler1}) implies averaging of the former over the quantum state. In the
functional integral formulation of the problem, classical equations
(\ref{Eq:Euler1}) correspond to the saddle-point treatment of the functional
integration. However, loop corrections can also contribute to the average
density and current and generate new terms in (\ref{Eq:Euler1}). Below we
discuss both aforementioned effects.

\subsection{Correction to Hamiltonian}
\label{s3.1}

Let us first explore corrections to the Hamiltonian 
(\ref{Eq:QuntumHamiltonian}). 
For a while  we put the loop corrections aside (we will return to them in
Sec.~\ref{s3.2}) and thus make no distinction 
between classical and quantum equations of motion. 

We start with  the Haldane's  theory\cite{haldane81} that accounts for a
discrete nature of particles as well  as for their spectrum
\cite{haldane-footnote}.
Within this model, the fermionic operator is represented by an infinite sum 
\begin{equation}
\label{Haldane}
\hat{\Psi}(x)=\sqrt{\hat{\rho}}e^{i\hat{\theta}(x)} \left(\sum_l e^{-il\pi\hat{\phi}(x)}\right)\,, 
\end{equation}  
where the  bosonic fields have the standard commutation relations 
\begin{equation}
[\hat{\theta}(x),\hat{\phi}(x')]=-\frac{i}{2}\rm{sgn}(x-x')\,,
\end{equation}
and are related to the velocity  and density fields as
\begin{eqnarray}
\hat{v}(x)=\partial_x\hat{\theta}(x)\,,\qquad
\hat{\rho}(x)=\partial_x\hat{\phi}(x)\,.
\end{eqnarray}
After substituting Eq.~(\ref{Haldane}) into the free Hamiltonian 
\begin{equation}
\hat{H}=-\frac{1}{2}\int dx \hat{\Psi}^+(x)\nabla^2\hat{\Psi}(x)\,,
\end{equation}
one obtains
\begin{equation}
\label{Ham_m}
H=\frac{1}{2}\int dx\sum_{l {\rm \,\, odd}} \bigg(\hat{\rho}  \hat{\theta}_x^2+\pi^2 l^2\hat{\rho}^3+\frac{1}{4}\frac{\hat{\rho}_x^2}{\hat{\rho}}\bigg)\,.
\end{equation}
This result contains an infinite summation over odd integers $l$, and formally diverges.
To properly define  this  series,  one needs to regularize the divergent sums.  
This can be achieved by   describing the  series as an expansion of an
analytic function of some argument 
($y(z)=\sum_{l} y_l z^l$).
A series of this type has to be summed within the range of its convergence and
then analytically continued to $z=1$. 
Bearing such a procedure in mind and comparing Eq.~(\ref{Ham_m})  with
Eq.~(\ref{Eq:QuntumHamiltonian}),
we establish that such regularization implies
\begin{eqnarray}
\sum_{l \rm \,\, odd} 1=1 \ \ \ \text{and} \ \ \  \sum_{l \rm \,\, odd}
l^2=1/3\,.
\end{eqnarray}
Thus, one  obtains
\begin{equation}
 \label{Ham_Haldane}
\hat{H}=\int dx\left(\frac{1}{2}\hat{\rho} \hat{v}^2 +\frac{\pi^2\hat{\rho}^3}{6}+\frac{1}{8}\frac{\hat{\rho}_x^2}{\hat{\rho}}\right)\,.
\end{equation}
The first two terms in Eq.~(\ref{Ham_Haldane}) are contained in 
the Hamiltonian (\ref{Ham_m}).  However, the last term in
Eq.~(\ref{Ham_Haldane})
represents the gradient corrections that are beyond the Hamiltonian
(\ref{Ham_m}).
Note that though  the derivation outlined above may not appear rigorous,
there is no ambiguity in determining the last term in 
Eq.~(\ref{Ham_Haldane}).  Indeed, the regularization procedure we use is fully
determined  by the first two terms in the Hamiltonian. Therefore, the
coefficient in front of  the third term is unambiguously determined. 
The Hamiltonian (\ref{Ham_Haldane}) leads to the following equations of motion
\begin{eqnarray}
\partial_t \hat{\rho}+\partial_x(\hat{\rho} \hat{v})=0\,, \qquad
\partial_t \hat{v}+\hat{v}\partial_x\hat{v}+\partial_x \hat{w}=0 \,.
\label{Eq:NoHilbert1}
\end{eqnarray}
Here 
\begin{eqnarray}
\hat{w}=\frac{\pi^2\hat{\rho}^2}{2}-\frac{1}{4}\partial_x^2\log\hat{\rho}-
\frac18\left(\partial_x\log\hat{\rho}\right)^2
\label{Eq:NoHilbert2}
\end{eqnarray}
is the enthalpy of the Fermi gas. The first term in Eq.(\ref{Eq:NoHilbert2}) is the 
pressure of a homogeneous Fermi gas, while the last two terms describe the
cyclotronic pressure that accounts for the finite density gradient.  
Interestingly enough, this latter contribution is  quite  universal and appears
also in the Madelung fluid\cite{stone} as well as 
in the hydrodynamic form of the Gross-Pitaevskii equation~\cite{hoefer06}. 
The presence of the finite gradient terms stabilizes the classical equation of
motions. Thus, it is natural to ask a question whether
Eqs.~(\ref{Eq:NoHilbert1}), (\ref{Eq:NoHilbert2}) are sufficient to describe
the evolution of the density pulse discussed in Sec.~\ref{s2.1}. 

\begin{figure}
 \includegraphics[width=230pt]{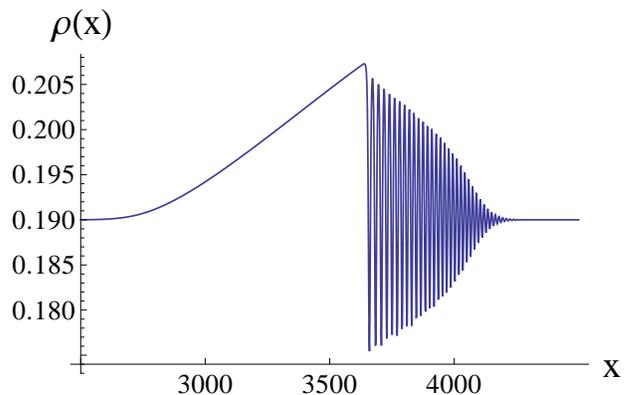}
\caption{\small Results of numerical solution of hydrodynamic equations
(\ref{Eq:NoHilbert1}), (\ref{Eq:NoHilbert2}). 
The initial density pulse was gaussian with the parameters used previously in
Sec.~\ref{s2}. The period of 
oscillations induced by the shock is determined by the competition
between dispersion and non-linearity, $\delta x\sim
\sqrt{\lambda_F/\Delta\rho}$. It is parametrically smaller then the one found
in Sec.~\ref{s2} within a direct analysis of the free-fermion problem.}  
\label{Fig:NoHilbert}
\end{figure}

To answer this question, we simulate the evolution of the density pulse (\ref{pulse_Gaussian}) 
in accordance with  Eq.~(\ref{Eq:NoHilbert2}). The results of this analysis
shown in Fig.~\ref{Fig:NoHilbert} clearly indicate the formation of a region of
oscillations in the density profile.
However, the period of the oscillations is parametrically different 
from that obtained from the direct quantum-mechanical solution of the
free-fermion problem of Sec.~\ref{s2}, see Appendix \ref{app_Haldane}
for details.  Indeed, the spatial scale of the oscillations is determined 
by the competition between the  non-linearity and the dispersion. 
In the present case, a simple estimate gives
\begin{equation}
\delta x\sim
\frac{1}{\sqrt{\rho_\infty\Delta\rho}}\sim\sqrt{\frac{\lambda_F}{\Delta\rho}}\,.
\label{period-KDV}
\end{equation}
This is smaller by a factor $(\lambda_F \Delta\rho)^{1/2}\ll 1$ than the result
(\ref{Eq:period}) of the direct solution of the free-fermion problem.
We thus conclude that equations (\ref{Eq:NoHilbert1}), (\ref{Eq:NoHilbert2})
yield a {\it parametrically wrong} scale for the density ripples: the
dispersive term  in Eqs.~(\ref{Ham_Haldane}), (\ref{Eq:NoHilbert2}) is
too weak. Thus, in our search for hydrodynamics  we must resort to loop
corrections to the equations of motion.  

Before passing to the analysis of loop corrections, let us make the
following comment. While the dispersive term in Eq.~(\ref{Ham_Haldane}) turns
out to be parametrically small in comparison to quantum effects for free
fermions, such a term (with a parametrically enhanced prefactor) will become a
dominant dispersive term for the case of electrons with finite range
interaction with a sufficiently large interaction radius, see Sec.~\ref{s4.1}.
Consequently, the semiclassical analysis of  Eq.~(\ref{Ham_Haldane}) (with an
appropriately modified coefficient of the last term) performed above and in
Appendix \ref{app_Haldane} will become a controllable description in that case,
as discussed in Sec.~\ref{s4.1}.

\subsection{Loop corrections}
\label{s3.2}

Let us  follow our phenomenological approach and try to guess the form of the
loop corrections  to the enthalpy (\ref{Eq:NoHilbert2})
on the basis of our knowledge of the characteristic scale of the ripples. It
is easy to see that to produce the correct period of the density oscillations
the correction   should scale as first power of momentum. A simple term of the
form $\partial_x \rho$ is not acceptable as it would break the symmetry with
respect to the spatial inversion. This symmetry can be saved however by
inclusion of the Hilbert transform $\widehat{H}$, 
\begin{equation}
 \delta w \sim \widehat{H}\partial_x \rho\,.
\label{Eq:Entalpy_Correction}
\end{equation}
By definition, in momentum domain the Hilbert transform acts according to
$ \widehat{H}\rho_k=-i\pi \sign(k) \rho_k$. 
From now on we will reserve a special notation $\widehat{A}_2$ for the operator $\widehat{H}\partial_x$. The reason for such a notation  will become clear in Sec.~\ref{s4}. In momentum space
\begin{equation}
 \widehat{A}_2\rho_k=\pi |k|\rho_k\,.
\label{Eq:A2}
\end{equation}

The enthalpy corrections of the form (\ref{Eq:Entalpy_Correction}) were first suggested by Jevicki~\cite{jevicki92}
 in his study of the $\rho^3$-theory defined by the the Hamiltonian
(\ref{Eq:QuntumHamiltonian}). He pointed out 
that such a theory contains two single-particle branches. In fermionic language
these correspond to the electron and hole parts of the spectrum,
\begin{equation}
 \epsilon_{p(h)}=k_F|k|\pm \frac{k^2}{2}\,,
 \label{spectrum-branches}
\end{equation}
with subscripts $p$ and $h$ referring to particles and holes, respectively.
Further, it was observed in Ref.~\onlinecite{jevicki92} that each of the
Lagrangians
\begin{eqnarray}
 L_{p(h)}=\int dx \left(\frac{1}{2}\frac{\phi_t^2}{\phi_x}-\frac{1}{8}\frac{\phi_{xx}^2}{\phi_x}-\frac{\pi^2}{6}\phi_x^3
\mp\frac{1}{2}
\phi_x\widehat{A}\phi_{x}\right),
\label{Eq:JevickiLagrangian}
\\
\phi=\partial_x^{-1} \rho\,.\quad
\end{eqnarray}
when treated semiclassically (i.e. on the saddle-point level), reproduces
correctly 
the dispersion relation for the corresponding branch of
excitations (\ref{spectrum-branches}). In other words,
Eq.~(\ref{Eq:JevickiLagrangian}) is an effective semiclassical theory that
takes into account explicitly quantum  corrections of the original cubic
theory (\ref{Eq:QuntumHamiltonian}). 

 The term with the Hilbert transform in Eq.~(\ref{Eq:JevickiLagrangian}) gives 
rise to $k^2$ correction to the linear spectrum of the conventional 
bosonization, see Eq.~(\ref{spectrum-branches}). It is known that loops
in the perturbative diagrammatic treatment (in the context of equilibrium
problems) of the Hamiltonian (\ref{Eq:QuntumHamiltonian})  lead indeed to such
an effect  \cite{samokhin98,aristov07}. Specifically, with loops taken into
account, the support of the bosonic spectral weight in the $(\omega, k)$-plain,
which is just a line $\omega=k_F k$ for the linear electronic spectrum, starts
to receive a finite width\cite{footnote-spectrum} of order $k^2$. We thus see,
that the inclusion of the term (\ref{Eq:Entalpy_Correction}) into the enthalpy
is a natural way to simulate the effect of loop corrections.  

\begin{figure}
 \includegraphics[width=245pt]{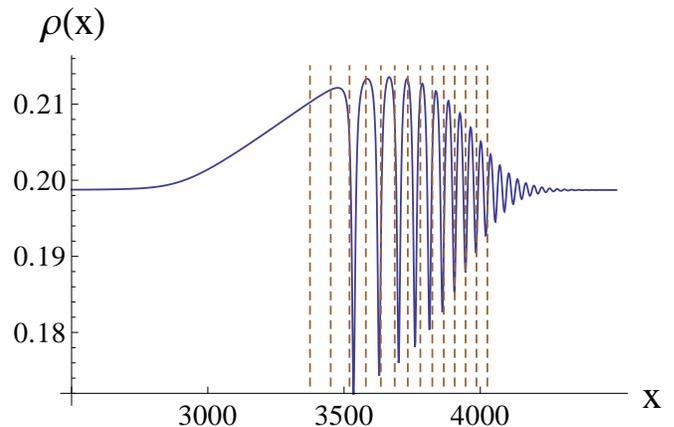}
\caption{\small Results of numerical solution of the semiclassical hydrodynamic
equations (\ref{Eq:NoHilbert1}), (\ref{Eq:EntalphyFinal}). 
Vertical lines mark the positions of density maxima in the exact
quantum-mechanical result for free fermions, see Sec.~\ref{s2}. }
\label{Fig:Hilbert}
\end{figure}

Motivated by this findings, we try to apply the
effective semiclassical Langrangians (\ref{Eq:JevickiLagrangian}) to our
problem. At this point, a question naturally arises: which of the
two Lagrangians $L_p$, $L_h$ should we choose [i.e., which sign should we choose
in Eq.~(\ref{Eq:JevickiLagrangian})]? We argue here in the following way. Let
us assume that the original density perturbation is positive, i.e, has a form of
a hump as shown in Fig.~\ref{Fig:Ur}. Such a perturbation can be obtained by
generating particle excitations on top of a homogeneous vacuum state. Therefore,
we choose the Lagrangian $L_p$ as appropriate in this situation. Similarly, in
the case of a dip-like (i.e, negative) density perturbation, the Lagrangian
$L_h$ should be taken. This choice is by no means innocent, as will be discussed
in more detail below.
%

For definiteness, we consider a hump-like excitation (as was also done in
Sec.~\ref{s2}) and thus the Lagrangian $L_p$. 
Corresponding hydrodynamic equations  (\ref{Eq:NoHilbert1}) with the enthalpy
\begin{equation}
 w=\frac{\pi^2\rho^2}{2}-\frac{1}{4}
\partial_x^2\log\rho-\frac18\left(\partial_x\log\rho\right)^2+\widehat{A}\rho
\label{Eq:EntalphyFinal}
\end{equation}
are of the Benjamin-Ono type. They were studied previously in the
literature in the context of Calogero model 
(see Sec.~\ref{s4.2}).  In Sec.~\ref{s3.3} we analyze these equations and
compare the outcome to the result of the fermionic solution presented in
Sec.~\ref{s2.2}.

\subsection{Non-local hydrodynamics of free fermions}
\label{s3.3}

As a first step of our analysis of the classical hydrodynamics defined by
Eqs.~(\ref{Eq:NoHilbert1}), (\ref{Eq:EntalphyFinal}), we have performed their 
numerical simulations for the initial density used previously in our quantum
computations. The result is shown in Fig.~\ref{Fig:Hilbert}. The dashed vertical
lines mark the positions of the maxima in the exact fermionic density analyzed
in Sec.~\ref{s2} (the same lines as in Fig.~\ref{Fig:WiegnerFunctionFinal}).
We observe a very good agreement between the free-fermion problem and the
classical hydrodynamics \ref{Eq:NoHilbert1}), (\ref{Eq:EntalphyFinal})
in the period of the oscillations induced by the shock. This agreement becomes
essentially perfect close to the front edge of the impulse. 

To explore analytically the oscillations emerging in the hydrodynamics
Eqs.~(\ref{Eq:NoHilbert1}), (\ref{Eq:EntalphyFinal}) at times exceeding the
shock time $t_c$, we employ the 
Whitham modulation theory \cite{whitham}. Within this approach, one
considers the solution to hydrodynamic equations in the shock region (the
interval between the points $x_l$ and $x_t$ of Fig.~\ref{Fig:pFBolzman} as
a periodic single-phase wave with slowly modulated parameters (wave vector,
frequency amplitude, etc.). The modulation equations for those parameters are
obtained from the Lagrangian averaged over a period of oscillations. For the
Lagrangian $L_p$ the single-phase periodic wave was found in
Ref.~\onlinecite{Polychronakos1995}:
\begin{equation}
\phi(x, t)=\rho_0 x-\gamma t +\Phi(\theta)\,, \qquad \theta=k x-\omega t\,.
\label{Eq:SinglePhaseWave}
\end{equation}
Here $\rho_0$ and $\gamma$ represent mean density and the current in the wave;
$k$ and $\omega$ are the wave vector and frequency, and $2\pi$-periodic function
$\Phi(\theta)$ is defined by its derivative
\begin{eqnarray}
 \dot{\Phi}=\frac{1}{2\pi}\left(1-\frac{\sinh a}{\cosh a-\cos\theta}\right)\,,
\label{Eq:Polychronakos1}
\\
  \tanh a=\frac{4\pi k^3\rho_1^3}{k^4\rho_1^2+4\pi^2k^2\rho_1^4-4(k\gamma-\omega\rho_0)^2}\,,
\label{Eq:Polychronakos2}
\\
\rho_1=\rho_0+\frac{k}{2\pi}\,.
\label{Eq:Polychronakos3}
\end{eqnarray}
The density in the wave $\rho=\partial_x\phi$ reads
\begin{equation}
 \rho=\rho_1-\frac{k}{2\pi}\frac{\sinh a}{\cosh a-\cos\theta}\,.
 \label{rho-BO}
\end{equation}
The parameter $a$ controls the amplitude of the periodic wave
\begin{equation}
 A=\frac{\rho_{\rm max}-\rho_{\rm min}}{2}=\frac{k}{2\pi\sinh a}\,,
\end{equation}
as well as its shape.
In the limit $a\gg 1$ Eq.~(\ref{rho-BO}) reduces to weak harmonic oscillations,
while in the opposite limit, $a\ll 1$, one gets a train of well separated
solitons, each of them carrying exactly one electron. 

The modulation equations for parameters $\gamma$, $\rho_0$, $k$ and $\omega$
were derived in Ref.~\onlinecite{gutman07} where a specific (Lorentzian) form of
the original pulse was used 
(see also Refs.~\onlinecite{matsuno98,matsuno-book} for the discussion of
modulation equations
for a closely related Benjamin-Ono equation). In Appendix
\ref{Ap:ModulationEq} we present a general derivation of modulation equations.
The result is conveniently presented in terms of four Riemann invariants 
$u_i$ satisfying
\begin{equation}
 \partial_t u_i +u_i\partial_x u_i=0\,, \qquad i=0\,,\ldots 4\,.
\label{Eq:Riemann}
\end{equation}
The parameters of the wave are given by
\begin{eqnarray}
 k=u_2-u_1\,,
\label{Eq:k}
\\
\omega=\frac{1}{2}\left(u_2^2-u_1^2\right)\,,
\label{Eq:omega}\\
\rho_0=\frac{u_3-u_2+u_1-u_0}{2\pi}\,,
\label{Eq:rho0}\\
\gamma=\frac{-u_0^2+u_1^2-u_2^2+u_3^2}{4\pi}\,.
\label{Eq:gamma}
\end{eqnarray}

The modulation equations should be supplemented by the boundary conditions at the ends of the shock region $x_t$ and $x_l$.
These conditions consist of the requirement that the average density and current
$\rho_0$ and $\gamma$ match those dictated by the Euler equation
(\ref{Eq:Euler}) in the regions without population inversion ($x<x_l$
and $x>x_t$). We thus have
\begin{eqnarray}
\rho_0 =\frac{p_F^{(3)}+p_\infty}{2\pi}\,, \quad \gamma=\frac{\left(p_F^{(3)}\right)^2-p_\infty^2}{4\pi}\,, \quad x=x_t\,,
\label{Eq:ModBoundary1}
\\
\rho_0 =\frac{p_F^{(1)}+p_\infty}{2\pi}\,, \quad
\gamma=\frac{\left(p_F^{(1)}\right)^2-p_\infty^2}{4\pi}\,, \quad x=x_l\,,
\label{Eq:ModBoundary2}
\end{eqnarray}
where we used the notation $p_F^{(i)}$ introduced for three branches of the
Fermi momentum in Sec.~\ref{s2}, see Fig.~\ref{Fig:pFBolzman}.
The solution of the equations for Riemann invariants with these boundary
conditions is given by
\begin{eqnarray}
 u_i=p_F^{(i)}\,,\qquad i=1,2,3 \,,
\label{Eq:ui}
\\
u_0=-p_\infty\,.
\end{eqnarray}

The modulation theory described above reveals a deep connection between the
hydrodynamic system (\ref{Eq:NoHilbert1}, \ref{Eq:EntalphyFinal}) and the free
fermions. Indeed, according to Eq.~(\ref{Eq:ui}), the Riemann invariants $u_i$,
$i=1,2,3$, characterizing hydrodynamic density oscillations in the shock region
are exactly equal to three branches $p_F^{(i)}$ of the Fermi surface of free
fermions in the population-inversion regime. Furthermore, the equipotential
lines of the action $S[X, p]$ that played a central role in our ``fermionic
analysis'' of the density ripples evolve according to exactly the same Euler
equation (\ref{Eq:Euler}) as that for Riemann invariants,
Eq.~(\ref{Eq:Riemann}). 

Equations (\ref{Eq:k}), (\ref{Eq:ui}) allow us to make a precise statement on
the period of oscillations:
\begin{equation}
 \delta x = \frac{2\pi}{p_F^{(2)} - p_F^{(1)}} \,.
 \label{period-BO}
\end{equation}
Thus, we see that close to the front  end of the pulse the period is
\begin{equation}
 \delta x=\frac{2}{\Delta\rho} \,,
 \label{period-BO-front}
\end{equation}
and coincides exactly with that of the density oscillations for the exact
quantum-mechanical solution of the free-fermion problem. It is easy to see that
near the top of the pulse the period is larger by a factor $\sim N^{1/3}$
(assuming for simplicity that the time $t-t_c$ that has passed after the shock
is of order $t_c$), again in agreement with the analysis of Sec.~\ref{s2.2}.
Equation (\ref{period-BO}) is fully consistent with the interpretation of
the oscillatory structure as Friedel oscillations between different
Fermi-momentum branches. 

The following comment is in order here. As has been explained above, when the
hydrodynamic theory (\ref{Eq:NoHilbert1}), (\ref{Eq:EntalphyFinal}) is
used to describe the behavior of free fermions, the choice of the particle
branch $L_p$ in Eq.~(\ref{Eq:JevickiLagrangian}) captures essential features of
the evolution of a density {\it hump}, while the hole-branch Lagrangian $L_h$ is
appropriate for a density {\it dip}. On the other hand, if we would try to
apply, e.g., $L_h$ for a density hump, it would fail completely. 
Specifically, it would predict the formation of the solitonic train in front
of the running pulse, i.e. a decomposition of the initial density perturbation
into well separated solitons (cf. Sec.~\ref{s4.2}), which never happens for free
fermions. It remains an open question whether there exists an improved
hydrodynamic theory that describes evolution of the free-fermion density
perturbation consisting of a combination of humps and dips. In this context, it
is also worth reminding the reader about the following. While the classical
hydrodynamics analyzed in this section perfectly reproduces the period of
free-fermion oscillations induced by a shock, it considerably overestimates
their amplitude. If there exists a better hydrodynamic description of this
problem, one might hope that it would be free also of this drawback.

\section{Interaction effects}
\label{s4}

In the previous sections we discussed the evolution of the
density perturbation in the free electron gas within (i) the 
exact ``fermionic'' approach and (ii) the phenomenological hydrodynamics.
We concluded that, in the latter formalism, quantum loop corrections are
crucial in determining the character of the dispersive regularization of the
shock. They can be modeled qualitatively by the non-local term 
(\ref{Eq:Entalpy_Correction}) in the enthalpy of free fermions. At this
level, the bosonized  Hamiltonian for free fermions (that corresponds to
the effective Lagrangian 
\ref{Eq:JevickiLagrangian}) takes the form (different for particle- and hole-like perturbations)
\begin{equation}
 H_{p(h)}=\int dx \left[\frac{1}{2}\rho v^2 +\frac{\pi^2\rho^3}{6}+\frac{1}{8}\frac{\rho_x^2}{\rho}\pm 
\frac12 \rho \widehat{A}_2\rho
\right]\,.
\label{Eq:H_boson_free}
\end{equation}
The term with $\rho_x^2$ coming from Haldane bosonization prescription is not
important in the low-gradient limit. 

In the present section we discuss 
modifications of the picture drawn above that arise due to the
electron-electron interaction. 
From the perspective of the ``fermionic'' solution of Sec.~\ref{s2.2}, one
obvious consequence of the interaction is 
the appearance of energy relaxation leading to local thermalization of the
distribution function. 
This thermalization will eventually wash out all the oscillating features of the
density. However, the corresponding time will be very large, since the lifetime
of electronic excitations in an interacting 1D system scales as a high power of
the mass $m$ (inverse curvature of the spectrum near the Fermi points), or
equivalently, of the Fermi momentum $p_F = mv_F$, see
Ref.~\onlinecite{imambekov11}
for a review. Specifically, at zero temperature the  lifetime of a quasiparticle
with momentum $k$ due to a long-range (smooth on the scale
$\lambda_F$) electron-electron interaction $V(r)$ is given by~\cite{khodas07,
imambekov11}
\begin{equation}
 \frac{1}{\tau_p}\sim
\left[V_0(V_0-V_{k-k_F})\right]^2\frac{(k-k_F)^4}{m^3v_F^6}\,,
\end{equation}
where $V_q$ is the Fourier transform of $V(r)$. 
We will be particularly interested below in the case of power-law decaying 
interactions
\begin{equation}
 V(r)=\frac{1}{m l_0^{2-\alpha}}\frac{1}{r^{\alpha}}\,,
\end{equation}
for which $V_0-V_q \propto q^{\alpha-1}$.  
Here, $1\leq \alpha<3$, and the length $l_0$ parameterizing the strength of the
interaction is the Bohr radius for the potential $V(r)$. Estimating
now the relevant momentum $k$ as $k-k_F\sim \Delta \rho$  we find the inelastic
decay rate\cite{footnote-cutoff}
\begin{equation}
 \frac{1}{\tau_p}\sim \frac{1}{m^7 v_F^6 l_0^8}
(l_0\rho_\infty)^{2\alpha-2}(l_0\Delta\rho)^{2+2\alpha}\,.
\label{ee-decay-rate}
\end{equation}
If the interaction falls off faster than $1/r^3$, one has $V_0-V_q \propto
q^2$; the corresponding result can be obtained by setting $\alpha=3$ in
Eq.~(\ref{ee-decay-rate}). On the other hand, the characteristic time scale for the density ripples
is the shock time $t_c\sim m\Delta x/\Delta\rho$. Assuming moderate interaction strength $l_0\sim 1/\rho_\infty$ 
we find
\begin{equation}
 \frac{t_c}{\tau_p}\sim N\left(\frac{\Delta \rho}{\rho_\infty}\right)^{2\alpha}\,.
\end{equation}
 We see that in the limit of small $\Delta\rho/\rho_\infty \ll 1$ the characteristic
time $\tau_p$ of inelastic decay given by Eq.~(\ref{ee-decay-rate}) is much
larger
than the shock time $t_c$. In other words, the relaxation effects  remain negligibly small at
times much larger than  $t_c$. In view of this, in the rest of the paper we
neglect the influence of inelastic relaxation on the dynamics and focus on other
interaction-induced effects that strongly affect the development of density
oscillations.

\subsection{Finite-range interaction}
\label{s4.1}

Let us first briefly discuss  the influence of finite-range interaction on the
dynamics of fermions. We parametrize 
the interaction potential at low momenta by the scattering length $l_0$ and
the effective interaction radius $l_{\rm int}$
\begin{equation}
 V_q=\frac{1}{m l_0}(1-q^2l_{\rm int}^2+\ldots) \simeq \frac{1}{ ml_0}-\frac{q^2
l_1}{m}\,,
\end{equation}
where $l_1 = l_{\rm int}^2 / l_0$.
 Correspondingly, the interaction-induced correction to the Hamiltonian takes the form
\begin{equation}
 H_{\rm int}=\frac{1}{2}\int dx \left[\frac{\rho^2}{l_0}-l_1 \left(\partial_x\rho\right)^2\right]
\end{equation}
Within the hydrodynamic description the fermionic mass $m$ manifests itself only via the time scale $t_c$ and we have 
set $m$ to unity (cf. the case of free fermions, Sec.~\ref{s2}).  
We see that the zero momentum  component of interaction gives rise to an
additional $\rho^2$ term in the Hamiltonian. The only effect of this correction
is  the renormalization of Fermi velocity. The $q^2$-part of the potential
$V(q)$ renormalizes the $\rho_x^2$ term in the free Hamiltonian. The resulting
term may compete with the last term (the one containing the Hilbert transform)
of Eq.~(\ref{Eq:H_boson_free}) in governing the dispersive regularization of
the shock dynamics. If the interaction range is not too long, $l_1 \ll
1/\Delta\rho$ (this is in particular the case for a short-range
interaction with $l_1 \lesssim \lambda_F$), the interaction-induced $\rho_x^2$
term can be discarded, and the dynamics will be the same as in the free-fermion
case. In the opposite limit of a very-long-range interaction, $l_1
\gg 1/\Delta\rho$, it is the the interaction-induced term that will control the
dispersive regularization.  Consequently, the period of
oscillations will not be given any more by the
free-fermion result (\ref{period-BO-front}) but rather will have a form of
Eq.~(\ref{period-KDV}) with $\lambda_F$ replaced by $l_1$, which yields
\begin{equation}
 \label{period-KDV-interaction}
 \delta x \sim \sqrt{l_1 / \Delta\rho}   \gg 1/\Delta\rho \,.
\end{equation}
Corresponding equations and their solutions  are
discussed in Appendix \ref{app_Haldane}; the only difference is that the
dispersive term is now enhanced by a factor $\sim \rho_\infty l_1$. Similarly to
what we will see below for power-law interactions (Sec.~\ref{s4.2} and
\ref{s4.3}), the character of resulting oscillations
will now depend on the sign of the initial pulse. Let us assume that the
interaction is repulsive. Then for an initial density dip the oscillation will
have a shape similar to those of free fermions [but with a larger period
according to Eq.~(\ref{period-KDV-interaction})]. On the other hand, for an
initial hump, the perturbation will decompose in a sequence of well-separated
solitons, cf. Fig.~\ref{Fig:Calogero} below.
The particle number  $q$ carried by each soliton is obtained from
Eq.~(\ref{soliton-charge-KDV}) by a replacement $1/\rho_\infty \to l_1$, which
results in  $q \sim \sqrt{l_1 \Delta_\rho} \gg 1$.


\subsection{Calogero model}
\label{s4.2}

The Calogero-Sutherland (CS) model\cite{calogero69,sutherland71} is a remarkable example of the quantum integrable model.
It appears in various branches of physics, such as spin chains, disordered metals and  fractional quantum Hall edges \cite{Forrester,Ha,Gangardt,Simon,wiegmann12}. 

In the CS model the particles  interact via an inverse-square potential 
\begin{equation}
 V(x)=\frac{\lambda(\lambda-1)}{m x^2}\,.
\end{equation}
Here $\lambda$ is the dimensionless interaction strength and $m$ is particle
mass. We will confine ourselves to the case of strong repulsion, 
$\lambda\gg1$. 

Being interested in the hydrodynamic description of the CS model
\cite{Awata,Polychronakos1995,Polychronakos_2006, bettelheim06,Stone_2008}, we
have to rewrite the CS Hamiltonian in terms of the particle density. While the
free part of the Hamiltonian after bosonization turns into the cubic Hamiltonian
(\ref{Eq:QuntumHamiltonian}), we need also a regularized expression for the
interaction term
\begin{equation}
 H_{\rm int}=\frac{1}{2}\int dx dx' V(x-x')\rho(x)\rho(x')\,.
\end{equation}
 The necessity of the regularization arises due to singularity of $V(x)$ at
$x=0$ and the corresponding ultraviolet divergence of the interaction 
at zero momentum. Taking the inverse particle density as the natural ultraviolet
cut-off in the problem,  we can rewrite the interaction term as
\begin{equation}
 H_{\rm int}\sim \lambda (\lambda-1)\int dx \rho^3-\frac{\lambda(\lambda-1)}{2}\int
dx \rho  \widehat{A}_2\rho \,.
\label{Eq:H_int_approx}
\end{equation}
The operator $\widehat{A}_2$ was defined in Eq.~(\ref{Eq:A2}).
The precise coefficient in front of the cubic term entering $H_{\rm int}$ is out of
control within this estimate. Also, terms with 
higher gradients  of the density may appear upon accurate regularization of the
model. A more rigorous  treatment of the CS model\cite{Polychronakos1995,
gutman07, stone08} leads  to a slight modification of (\ref{Eq:H_int_approx})
and results in
\begin{multline}
 H=\int dx\left[ \frac{\rho v^2}{2}+\frac{\pi^2 \lambda^2\rho^3}{6}
-\frac{\lambda(\lambda-1)}{2}\rho \widehat{A}_2\rho\right.\\\left.+\frac{
(\lambda-1)^2}{8}
\frac{\rho_x^2}{\rho}\right]\,.
\label{Eq:H_Calogero_Exact}
\end{multline}

The characteristic feature of the Calogero model is the scaling of the
interaction with distance  
which coincides exactly with that of the kinetic energy. At large coupling
constant, $\lambda\gg 1$, the potential energy dominates over the kinetic energy
at all scales and drives the model towards the semiclassical limit. Indeed, 
rescaling by  $\lambda$  the density and the space-time coordinates and
switching to Lagrangian formalism, one finds the action corresponding to
Eq.~(\ref{Eq:H_Calogero_Exact}):
\begin{equation}
 S=\lambda \int dx
dt\left[\frac{1}{2}\frac{\phi_t^2}{\phi_x}-\frac{1}{8}\frac{\phi_{xx}^2}
{\phi_x} -\frac{\pi^2}{6}\phi_x^3
+\frac{1}{2} \phi_x\widehat{A}_2\phi_{x}\right]\,.
\label{Eq:S_Calogero}
\end{equation}
Here $\partial_x\phi=\rho$. The large factor $\lambda$ in (\ref{Eq:S_Calogero})
justifies now the semiclassical approach.

\begin{figure}
 \includegraphics[width=230pt]{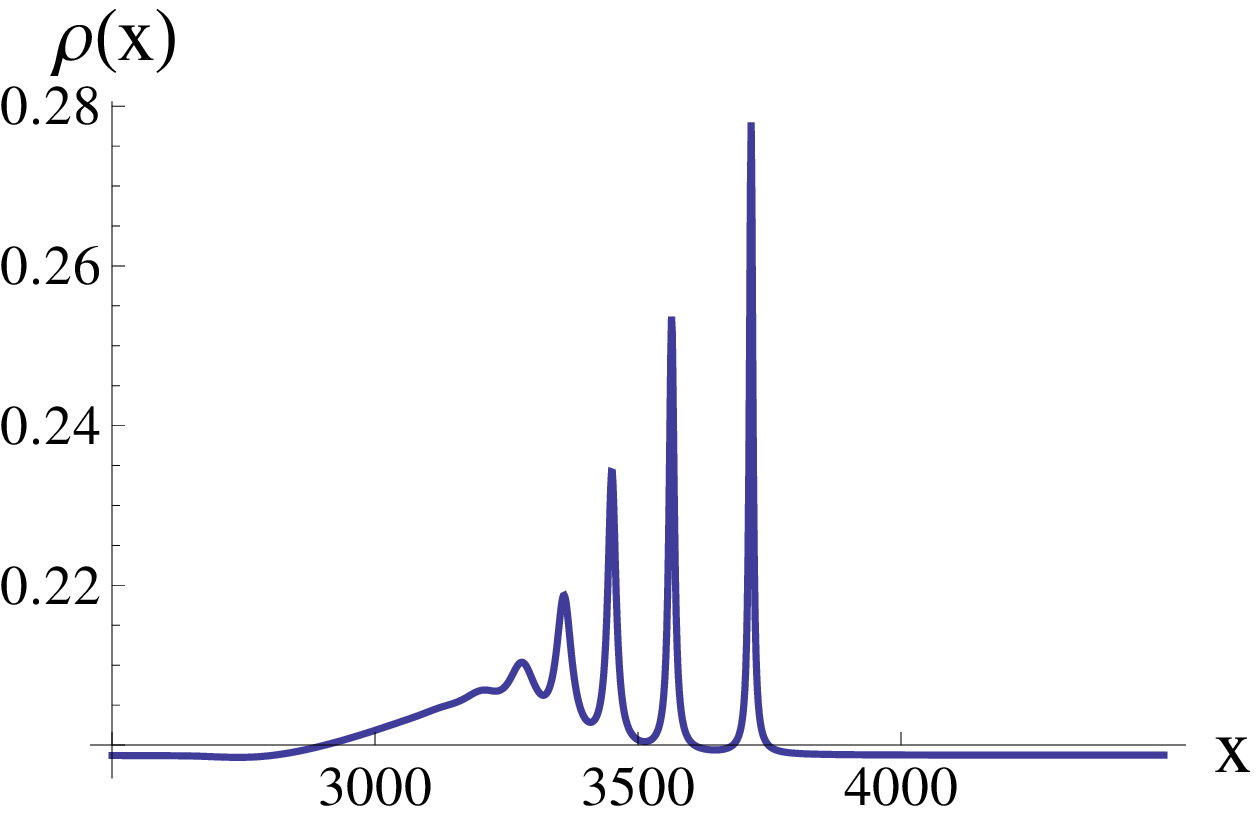}
\includegraphics[width=230pt]{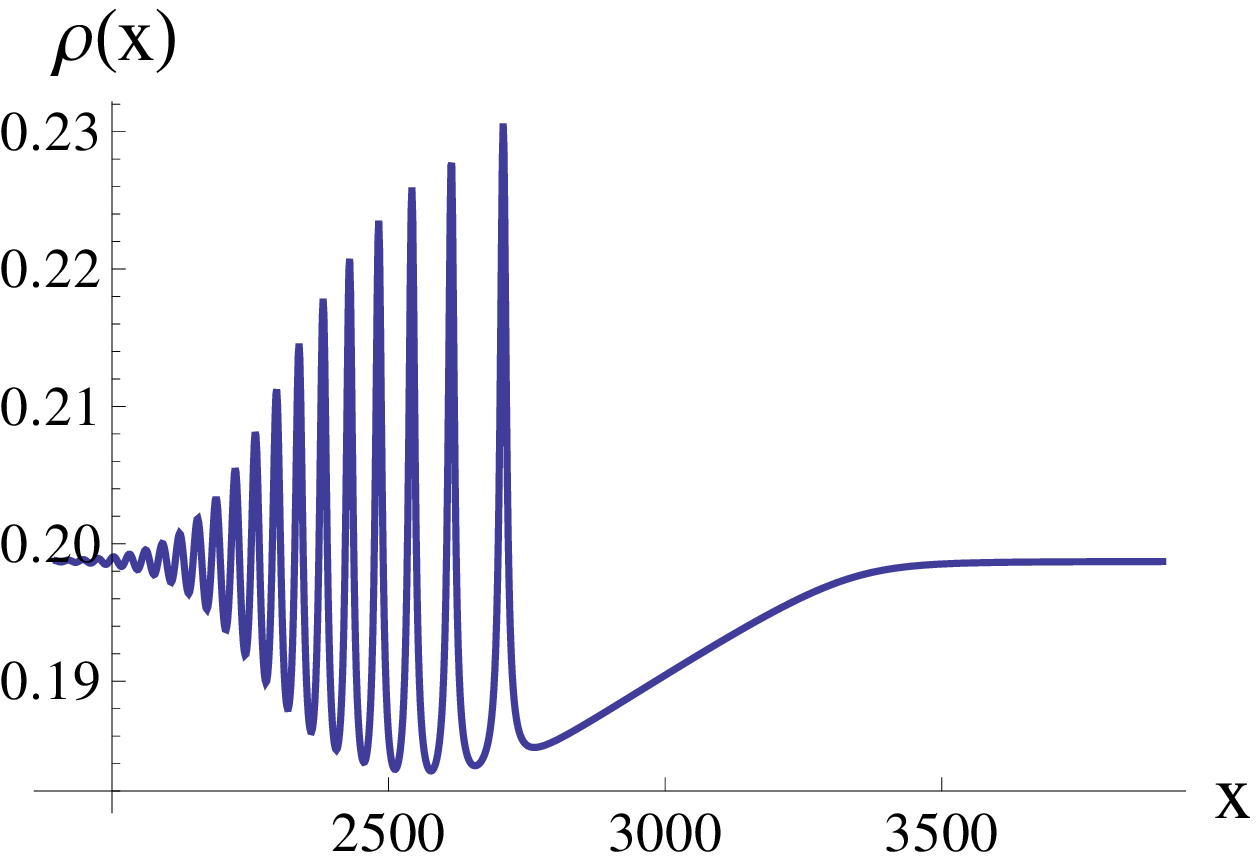}
\caption{\small Density profile for the Calogero model with strong repulsive
interaction after the shock, $t\simeq 5t_c$. Initial perturbation was a density
hump ({\it top panel}) and density dip ({\it bottom panel}).
In the case of a density hump, the initial gaussian density perturbation decays
into solitons carrying exactly one electron each. On the other hand, for a
downward density pulse in the initial state, nearly sinusoidal oscillations
develop in the shock region, so that the density evolution is similar to that of
free fermions. Note that for a density dip the shock occurs on the rear side of
the pulse. }
\label{Fig:Calogero}
\end{figure}

Note that the Lagrangian in Eq.~(\ref{Eq:S_Calogero}) is precisely the ``hole''
(not particle!) Lagrangian $L_h$  we  encountered in our discussion of  loop
corrections to hydrodynamics of free fermions [see
Eq.~(\ref{Eq:JevickiLagrangian})
of Sec.~\ref{s3.3}]. The corresponding equations of motion are the Euler
equations (\ref{Eq:NoHilbert1}) with the enthalpy
given by (\ref{Eq:EntalphyFinal}) except for additional minus sign in front of
the non-local term.  

This change of sign has a dramatic effect on the density evolution in the
system after the shock, as illustrated in Fig.~\ref{Fig:Calogero}
(top panel) where we plot the fermionic density at $t\approx 5 t_c$ for the same
initial density hump as was used previously. 
In the shock region, instead of the dispersive wave seen in
Fig.~\ref{Fig:Hilbert}, one observes the formation of a solitonic train. 
Thus, at late stages of the evolution the initial hump decays into well
separated solitons. Each of the solitons carries exactly one particle. The
quantization of solitonic charge, which is equal to unity,  is a
distinct feature of the strongly repulsive Calogero model. 

 The solitonic train in the shock region can be studied analytically via the solution 
of modulation equations discussed previously in Sec.~\ref{s3.3}.   One finds (see Appendix~\ref{Ap:ModulationEqS} for details) that close to the front edge of the train the height and width of the solitons (which are of Lorentzian shape) are given by
\begin{equation}
 \delta\rho=2\Delta\rho\,,\qquad \delta x=\frac{1}{2\pi \Delta\rho}\,.
\end{equation}

The density evolution is very much different (and much more similar to that
of free fermions) for the case of an initial density dip. The corresponding
data are shown in the bottom panel of Fig.~\ref{Fig:Calogero}; 
they are fully analogous to the previous results of Fig.~\ref{Fig:Hilbert}.
(Note that for a density dip the shock occurs on the rear
side of the pulse.) We see that a nearly sinusoidal dispersive wave is formed in
the shock region. 

Such a dramatic difference in the behavior of particle-like and hole-like
pulses has a simple qualitative explanation. A soliton can be formed when the
effects of non-linearity and dispersion on the velocity counteract; their
balance yields a soliton that moves preserving its shape. In the case of the 
Lagrangian (\ref{Eq:S_Calogero}) [which is equivalent, up to an overall factor
$\lambda$, to  $L_h$  of Eq.~(\ref{Eq:JevickiLagrangian})] the dispersive term
reduces the velocity, see Eq.~(\ref{spectrum-branches}) with the lower sign
corresponding to $L_h$. Therefore, solitons can form if the non-linear term
will enhance the velocity. This is the case when $\Delta\rho$ is positive, i.e.
for a density hump. 




\subsection{Coulomb and other slowly decaying interactions}
\label{s4.3}

\begin{figure*}
 \includegraphics[width=230pt]{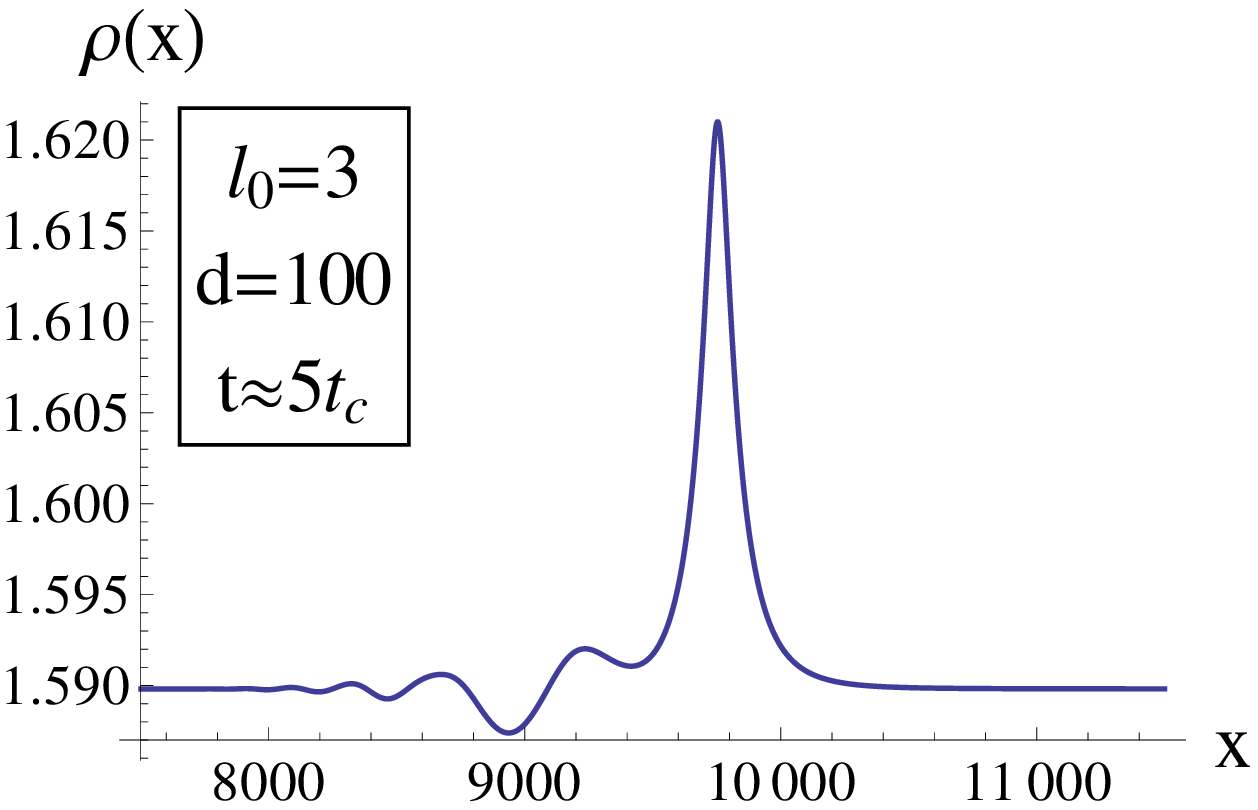}
 \includegraphics[width=230pt]{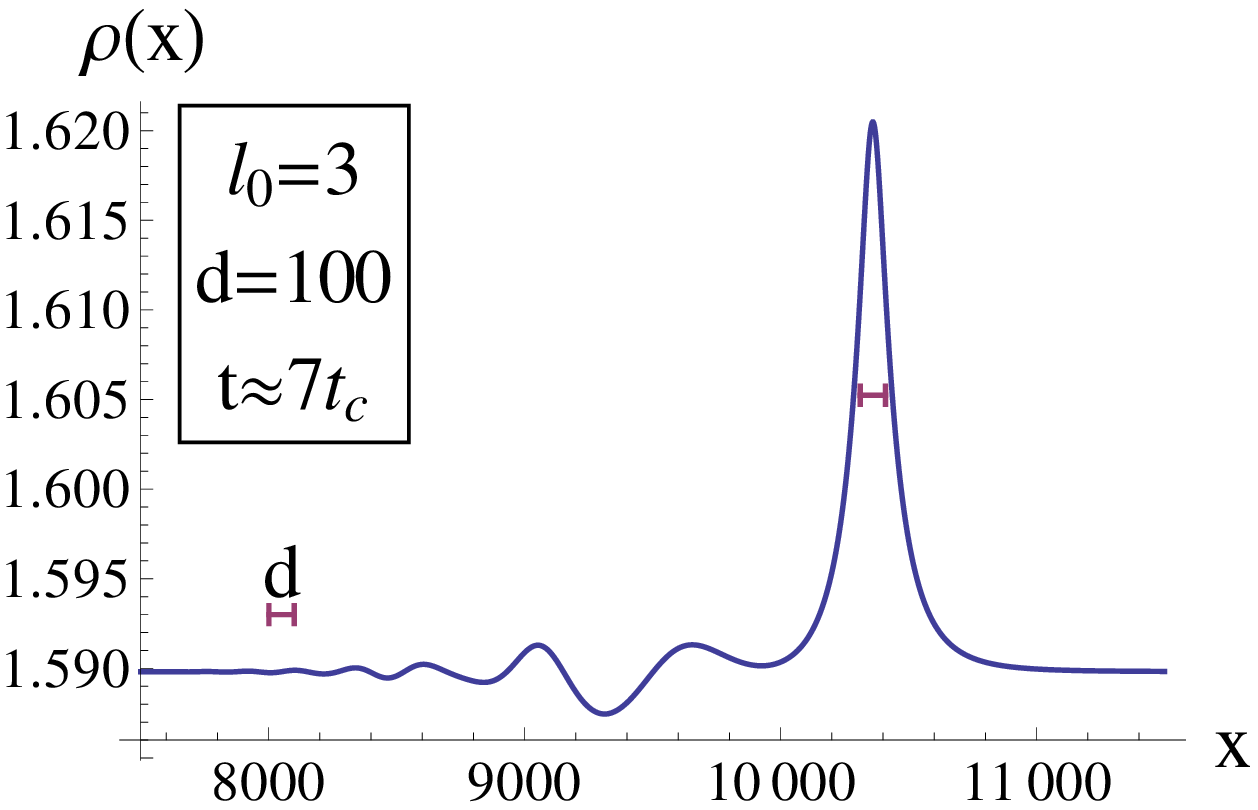} 
 \includegraphics[width=230pt]{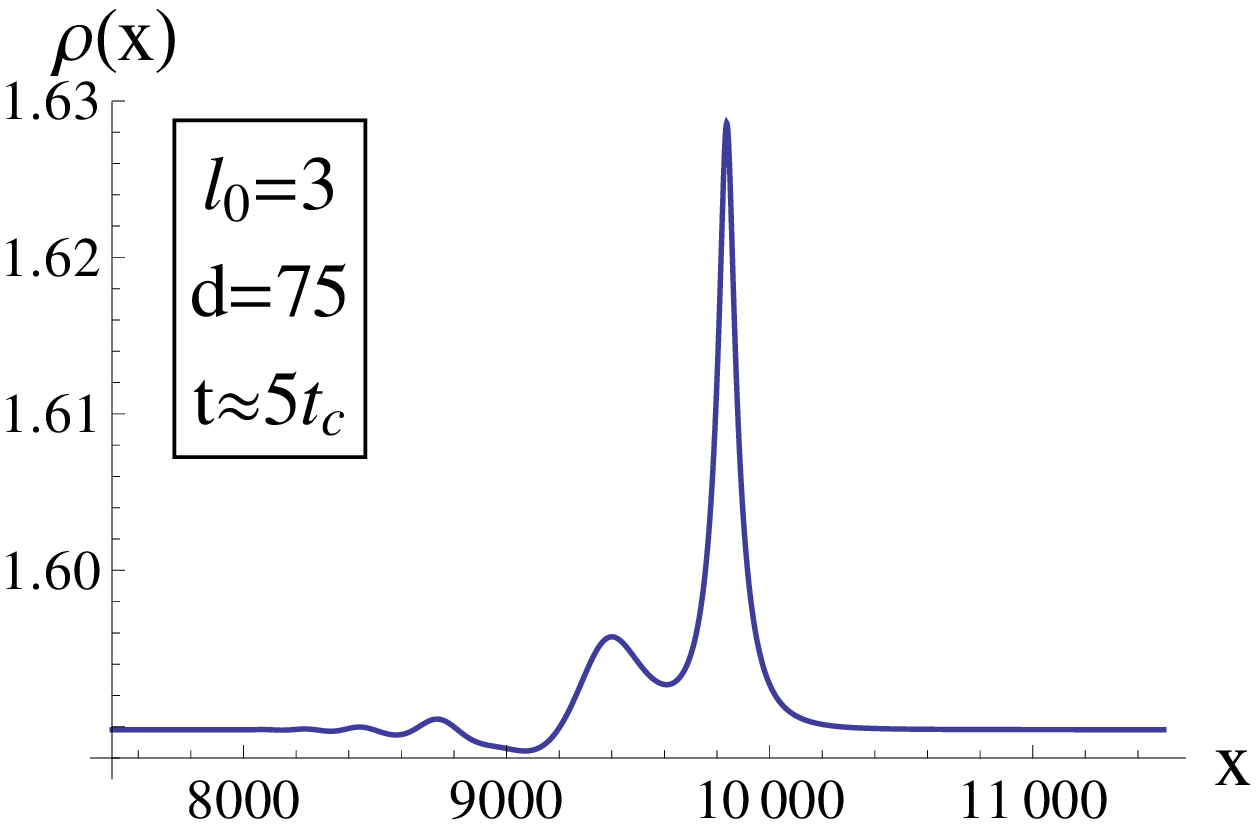}
 \includegraphics[width=230pt]{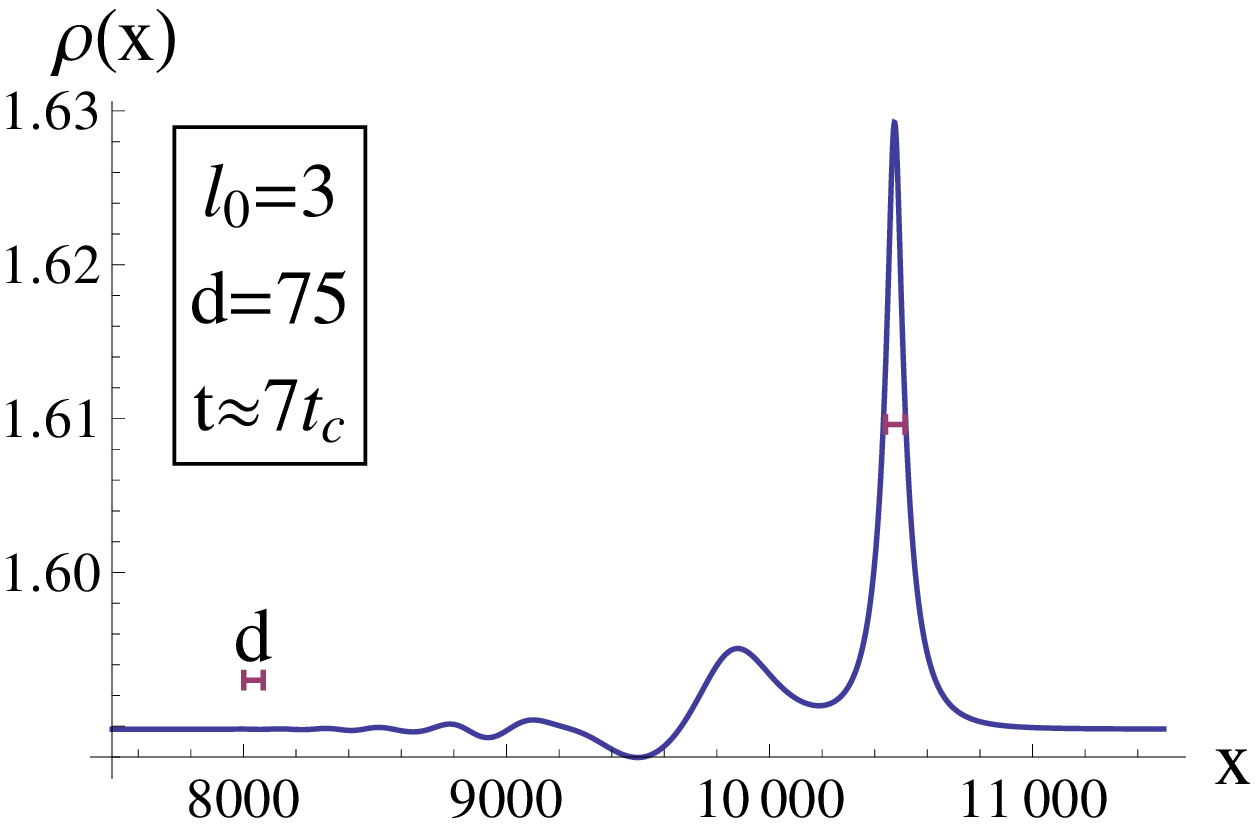} 
 \includegraphics[width=230pt]{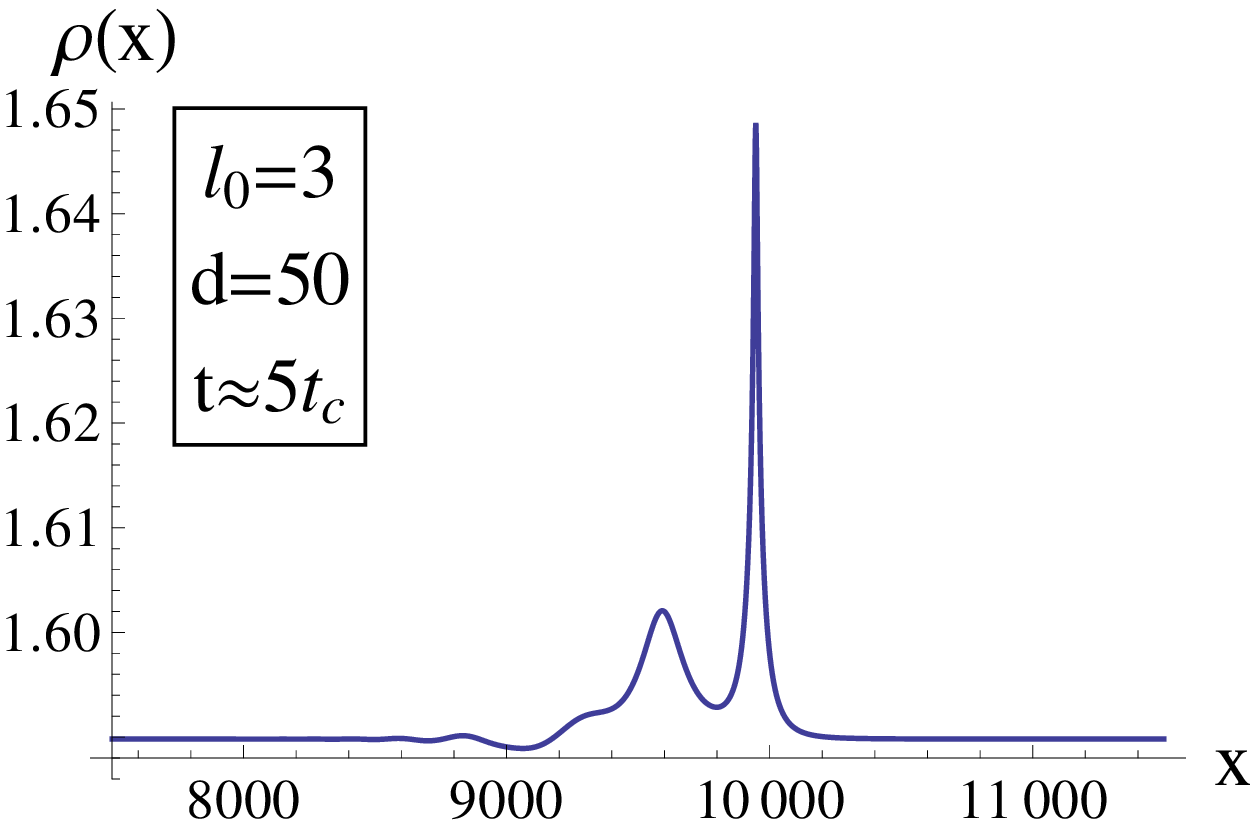}
 \includegraphics[width=230pt]{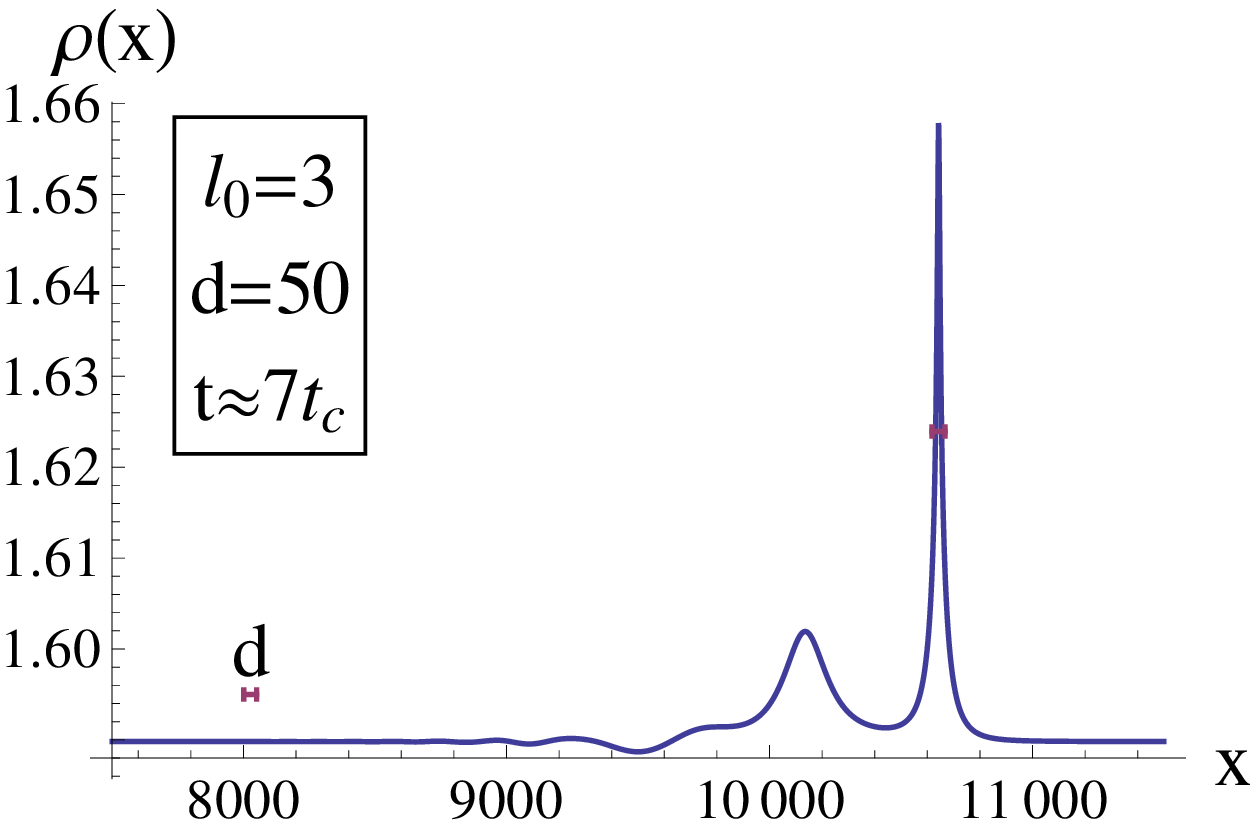} 
\caption{\small
Density profile after the shock for the case of Coulomb interaction.
The initial
density perturbation is positive and coincides with the one used in
Sec.~\ref{s2} (up to the constant equilibrium density $\rho_\infty$). 
The Bohr radius $l_0$, the screening length $d$, and the snapshot time $t$ are
indicated in each graph. For convenience of presentation, a constant coordinate
shift has been made: $x \to x - 20000$ for the left ($5t_c$) plots and  $x \to x
- 30000$ for the right ($7t_c$) plots.
For the upper two and the middle two graphs,  the 
half-maximum width of the main peak is approximately given by $d$, as shown in
the right plots.  For the two bottom graphs the width is smaller than $d$ by a
factor $\approx 1.5$.  This deviation from scaling with decreasing $d$ is
probably related to the fact that the condition (\ref{condition-l0}) becomes
less well satisfied.
}
\label{Fig:Coulomb}
\end{figure*}

\begin{figure}
 \includegraphics[width=230pt]{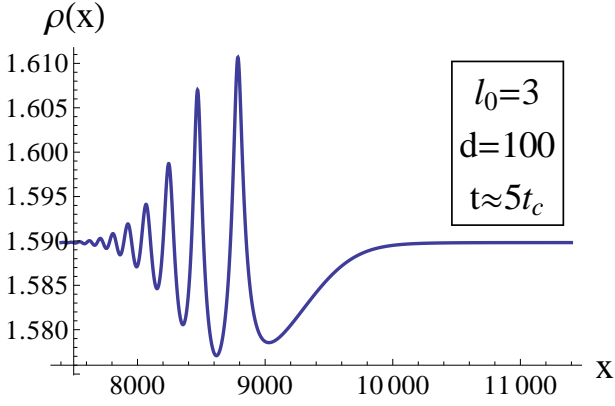}
 \caption{\small
 Density profile after the shock for the case of Coulomb interaction.
  The initial density
perturbation differs from the gaussian pulse of Sec.~\ref{s2} by the change of
sign. The legend indicates the  Bohr radius $l_0$ and the screening length $d$.
For convenience of presentation, a constant coordinate
shift $x \to x - 30000$ has been made.
 In contrast to the case of a positive (upward) density
pulse (Fig.~\ref{Fig:Coulomb}), a nearly sinusoidal oscillatory
behavior is observed in the shock region.}
\label{Fig:CoulombDip}
\end{figure}

\begin{figure}
 \includegraphics[width=230pt]{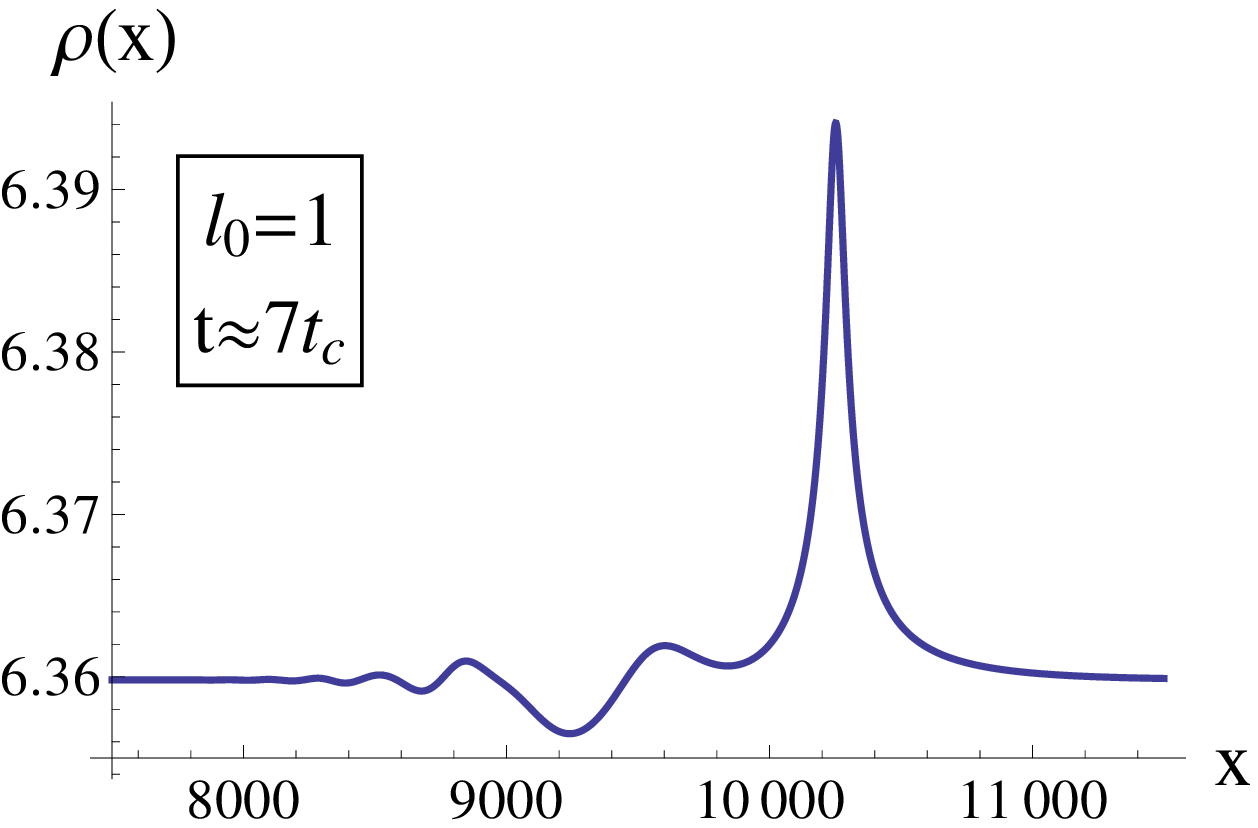}
 \includegraphics[width=230pt]{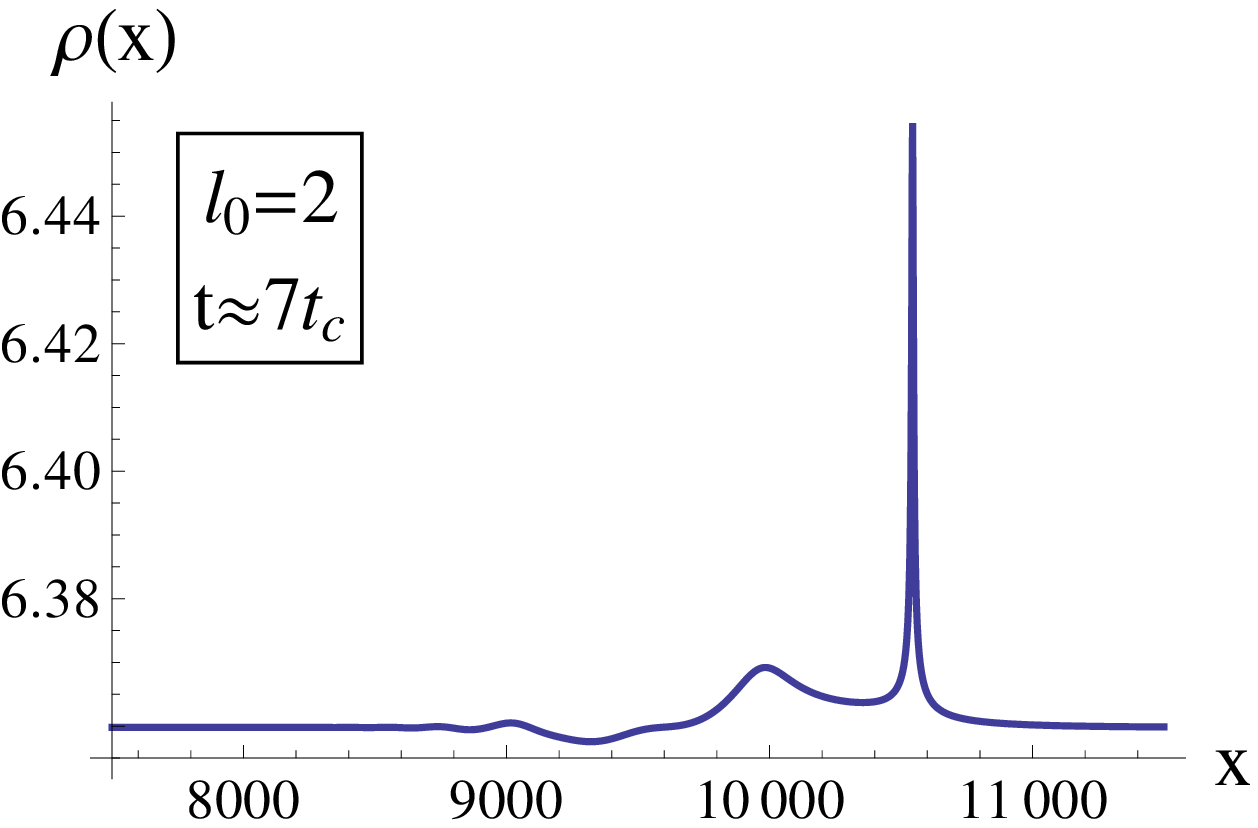} 
 \caption{\small Density  profile after the shock (time $t \simeq 7t_c$) for
electrons interacting via $r^{-3/2}$ potential. The initial density perturbation
is positive and
coincides with the one used in Sec.~\ref{s2} (up to the equilibrium density
$\rho_\infty$). The Bohr radius $l_0$ is indicated in the plots. 
For convenience of presentation, a constant coordinate
shift $x \to x - 120000$ has been made.
 }
\label{Fig:Power}
\end{figure}

Let us now turn to interactions decaying slower than the inverse distance
squared (for definiteness, we will assume a repulsive interaction),
\begin{equation}
 V_\alpha(r)=\frac{1}{m l_0^{2-\alpha}}\frac{1}{r^{\alpha}}\,,
\label{Eq:V_alpha}
\end{equation}
with an exponent $\alpha$ satisfying $1\leq \alpha<2$. The case
$\alpha=1$ corresponds to the Coulomb interaction and is the most relevant from
the experimental point view.
Throughout this section we will assume the interaction to be weak in the sense
that the parameter $r_s=\lambda_F/l_0$ is small, $r_s \ll 1$. (There is no
problem in analyzing the strong interaction regime, $r_s \gtrsim 1$, in a
similar way, and we expect a qualitatively similar behavior.)
For $\alpha=1$ we will also assume that the Coulomb interaction is screened at
a sufficiently large distance $d \gg l_0$.

The reasoning of the previous section which led us to Eq.
(\ref{Eq:H_int_approx}) is easy to generalize  for the present case with the
result 
\begin{eqnarray}
 H_{{\rm int}, \alpha} \simeq \frac{1}{l_0^{2-\alpha}}
\int dx \left(\rho^{\alpha+1}-\frac{1}{2 }\rho \widehat{A}_{\alpha} \rho\right)\,,\\
A_{\alpha}(q)=-2\Gamma[1-\alpha]\sin\frac{\pi\alpha}{2}|q|^{\alpha-1}\,, \qquad
\alpha>1 \,,\\
A_{1}(q)=\log q d\,, \qquad qd\gg 1 \,.
\end{eqnarray}
In the Coulomb case the $\rho^{\alpha+1}$ term in $H_{{\rm int}, \alpha}$ should
be replaced by $(\rho^2\log d \rho)/l_0$. 
For $\alpha > 1$, the precise numerical coefficient in front of the
$\rho^{\alpha+1}$ term can not be found within this reasoning.  
On the other hand, this term is small in parameter $r_s$ compared to the cubic
term in the Hamiltonian of free fermions which provides the dominant
nonlinearity. The only effect of the non-linear contribution $H_{\rm int}$ is a 
renormalization of Fermi velocity (small at $r_s \ll 1$) and we can omit it.
Combining the free bosonised Hamiltonian with the relevant (dispersive) part of
the interaction correction, we find
\begin{equation}
 H=\int dx\left[ \frac{\rho v^2}{2}+\frac{\pi^2 \rho^3}{6} -\frac12\rho
\widehat{A}_\alpha\rho\right]\,.
\label{Eq:H_alpha}
\end{equation}

As we have found previously, the loop contribution 
to the  equations of motion can be modeled by a correction to the Hamiltonian
of the form
\begin{equation}
 \delta H_{\rm loop} \simeq \int dx \rho \widehat{A}_2\rho\,.
 \label{loop-correction}
\end{equation}
Comparing the interaction-induced dispersive contribution [last term
in Eq.~(\ref{Eq:H_alpha})] to Eq.~(\ref{loop-correction}),
we see that the interaction controls the dispersive effects 
at scales larger than $l_0$.

The characteristic scale developed by the density perturbation after the shock
results from the trade-off  between 
non-linearity and dispersion. For $\alpha> 1$ the corresponding estimate
yields  the scale
\begin{equation}
 \delta x\sim l_0 \frac{1}{(l_0 \Delta \rho)^\beta}\,, \qquad
\beta=\frac{1}{\alpha-1}\,.
\end{equation}
We see that $\delta x$ is indeed much larger than $l_0$ provided that
\begin{equation}
 \label{condition-l0}
 l_0\,\Delta\rho \ll 1 \,.
\end{equation}
In the Coulomb case under the same assumptions we get 
\begin{equation}
\label{delta-x-coulomb}
\delta x\sim d \gg l_0 \,.  
\end{equation}
Thus, under the assumption (\ref{condition-l0}) the physics of
density oscillations is indeed dominated by scales much larger than $l_0$, so
that the neglect of loop corrections is justified.


Experience gained in the analysis of the Calogero model and free fermions allows
us to predict qualitative features of the density evolution, most prominently,
its dependence on the sign of the density perturbation and the sign of the
interaction. Specifically, we expect that for repulsive interaction and positive
perturbation a train of solitary waves should emerge in front of the pulse. 
Each solitary wave is expected to carry $\delta x\Delta \rho \gg 1$ particles.
On the other hand, a downward density perturbation (for the same repulsive interaction) 
will lead to formation of a nearly sinusoidal dispersive wave in the shock
region. The change of the sign of interaction will result in the interchange of
these two types of behavior.

To support the qualitative analysis presented above,  we have performed
numerical simulations of the hydrodynamic equations dictated by the Hamiltonian
(\ref{Eq:H_alpha}). Let us discuss the Coulomb case first.
Figure \ref{Fig:Coulomb} shows the density perturbation for times $t=5t_c$ and
$t=7t_c$ for electrons interacting via a Coulomb potential. The initial
density hump was Gaussian with the same parameters as in the previous sections
except for the equilibrium density 
$\rho_\infty$ which was taken larger  to ensure that $l_0\gtrsim
\lambda_F$~\cite{Remark1}. Values of the parameters $l_0$
and $d$ as well as of the time $t$  are indicated in each
of the plots. We clearly observe formation of a solitary wave and beginning of
the formation of a second one (better pronounced for smaller $d$).
According to the estimates presented above, in the
Coulomb case  the characteristic scale of the density oscillations emerging
after the shock should be given simply by the screening length $d$. 
This is indeed confirmed by our numerics. In particular, the half-maximum width
of the main peak in the plots with $d=100$ and $d=75$ is equal to $d$. The two
bottom plots (with the smallest $d$ equal to 50) demonstrate some deviation
from this scaling. This is possibly related to the fact that the condition
(\ref{condition-l0}) becomes less well satisfied in view of increasing
amplitude $\Delta\rho$ of the peak.

%

Figure~\ref{Fig:CoulombDip} illustrates the change of the density behavior upon
the change in of the sign of the density perturbation. As expected, for negative
$\Delta\rho$  we observe onset of nearly sinusoidal oscillations with a
period $\sim d$ in the shock region.

We have also performed numerical study of fermions interacting via the
intermediate potential $V_{3/2}(x)$.  The results 
are exemplified in Fig.~\ref{Fig:Power}. We observe that the density develops a
solitary wave, similarly to the case of Coulomb interaction
Fig.~\ref{Fig:Coulomb}. The scaling of the width of the soliton agrees well
with our above estimate of the characteristic scale for $\alpha=3/2$, 
\begin{equation}
 \delta x\sim \frac{l_0}{(l_0\Delta\rho)^2}\,,
\label{Eq:DeltaX32}
\end{equation}
if we use for $\Delta \rho$ the actual amplitude of the peak. (While in the
Calogero $1/r^2$ case the soliton amplitude $\Delta\rho$ is determined by that
of the initial pulse, this is no more true for $\alpha < 2$.) Note that the
parameter $l_0\Delta\rho$ remains sufficiently small ($\simeq 0.04$ for the
upper plot and 0.2 for the lower plot), so that the neglect of loop corrections
is reasonably well justified. 

 
The analytical arguments and numerical data presented above unambiguously show
that a sufficiently strong and sufficiently long-ranged interaction
dominates over quantum corrections in controlling the dispersive effects. In
this limit the Hamiltonian (\ref{Eq:H_alpha}) and corresponding hydrodynamic
equations provide a controlled description of the non-equilibrium dynamics in a
quantum many-body system.

\section{Summary and Outlook}
\label{s5}

In this article, we have explored the evolution of a density pulse in a
1D fermionic fluid. Our focus was on the regime of a wave ``overturn''
(population inversion) that is induced by spectral curvature. We showed that
beyond the corresponding time the density profile develops strong oscillations
with a period much larger than the Fermi wave length and performed a detailed
analysis of these oscillations. We have considered the case of free fermions as
well as various interacting models, including a finite-range interaction, CS
model, 
generic power-law interaction,  and screened Coulomb interaction. Our key
results can be summarized as follows:

\begin{enumerate}
 
 \item 
 For the case of free fermions we  have studied the problem by means of direct quantum simulations. Further, we have obtained analytical solution   using
the phase space representation and the  Wigner function.
The Wigner function of the initial state exhibits oscillations in the phase
space (as a function of the momentum). When the initial perturbation is allowed
to propagate (i.e. after the quench), the curvature of single particle spectrum
leads to the formation of inverse population of electrons at times
$t>t_c$. In this regime,
the oscillations of Wigner function in phase space induce real space density
oscillations, with each ``ripple'' containing a fraction of an electron. The
characteristic period of these oscillations is controlled by the amplitude
$\Delta\rho$ of perturbation and is independent of the equilibrium density
$\rho_\infty$ (or, equivalently, of the wave length $\lambda_F$). 
 
 \item
 We have also addressed  the free-fermion problem using a hydrodynamic approach.
The semiclassic equation of motion leads to formation of a shock in the regime
where the inverse population of fermions in the momentum space is generated.
This shock is regularized by gradient corrections to the Hamiltonian and
by quantum fluctuations. We show that for free fermions the latter effect is
more important. We model the  quantum correction by including in the theory a
dispersive term corresponding to a particle or hole branch of the fermionic
spectrum, depending on the sign of the initial perturbation. This yields two
different hydrodynamic theories (with a difference in the sign of the
dispersive term) for upward and downward density pulses. 

We show that this approach correctly captures the period of shock-induced
density oscillation but overestimates their amplitude. In the hydrodynamic
language, the formation of oscillations is caused by an interplay of  
the non-linearity (caused by the spectrum curvature)  and the dispersion
(dominated, in the case of free fermions, by quantum corrections carrying
information about the spectral curvature).
 
\item
The electron interaction leads to additional dispersive terms in the hydrodynamic equations.
For interaction that decays with the distance $r$ slower than $1/r^2$, such
terms dominate the long-distance (small momentum) behavior, and quantum
correction can be neglected. In this case the applicability of semiclassical
hydrodynamic equations becomes fully justified.  
The case of CS model ($1/r^2$ interaction) is marginal; the interaction-induced
dispersive term is dominant (and thus the semiclassical hydrodynamic approach is
fully controlled) if the interaction is strong, $\lambda \gg 1$. 

For the case
of a finite-range interaction the dominant dispersive term is provided by the
interaction only if the interaction radius is very big; otherwise, the
free-fermion results apply. 

\item
In the situations when the interaction controls the dispersive effects (and
thus the semiclassical hydrodynamic approach is fully under control), the impact
of interaction depends on its sign and the sign of the density
perturbation. 
Specifically, for a repulsive interaction and  a density dip 
(as well as for an attractive interaction and a density hump), 
we observe formation of nearly sinusoidal oscillatory structure similar to the
free fermions case. Quantitative characteristics of the oscillations (wave
length and a number of particle in each ``ripple'') are however in general
parametrically different compared to the free-fermion model.

On the other hand, for a repulsive interaction and a density hump 
(as well as for an attracting interaction and a density  dip),  the interaction
leads to the formation of a train of solitary waves. In general, the charge
(particle number) carried by each soliton is non-universal (depends on the type
and the strength of the interaction, and on the amplitude of the perturbation).
A notable exception is the CS model with $\lambda\gg 1$, when the solitons
carry a unit charge. 

\end{enumerate}

We hope that our predictions can be verified experimentally. There is a number
of electronic realizations of 1D fermionic systems, including carbon nanotubes,
semiconductor and metallic nanowires, as well as quantum Hall  and topological
insulator (quantum spin Hall) edges. For these electronic liquids, a model with
Coulomb interaction is expected to be applicable (except if special efforts are
made to strongly screen it). An alternative physical realization is provided by
systems of cold fermionic atoms. This is probably the most natural
experimental realization of the models of free fermions and of finite-range
interaction. 

Before closing, we list some of directions of further theoretical
research opened by the present paper; a work in some of these directions is
currently underway. 

\begin{enumerate}

\item An interesting question is whether it is possible to formulate a more
general classical hydrodynamic theory for free fermions that would
controllably capture evolution of a generic density perturbation, including both
upward and downward density pulses. 
Such a theory can be useful from the fundamental point of view, as 
well as for the problem in which quantum corrections and interaction effects are comparable.

\item  For models with power-law interaction other than CS model (including the
experimentally most relevant case of the Coulomb interaction), it is important
to complete analytical investigation of the emerging oscillations and
solitary waves and to explore the integrability of these theories. 

\item An important task is to perform ab initio calculations for many-body
quantum interacting system. The results should allow one to verify our
above predictions (obtained in the framework of the hydrodynamic theory) and
to explore the interplay of quantum corrections and interaction (e.g. in the
model with a finite-range interaction).  

\item Our results on evolution of a density perturbation should be also relevant
to strongly repulsive 1D bosonic problems, in particular, in view of the
equivalence between the Tonks-Girardeau gas and free fermions. It would be very
interesting to study the crossover from the quasi-condensate regime
characteristic for weakly interacting bosons \cite{hoefer06,kulkarni12} to the
Fermi-like behavior for
strong repulsion. On the experimental side, such a setup can be realized in
the framework of cold bosonic atoms.
 


\end{enumerate}

\section{Acknowledgments}

We thank M. Hoefer for sharing his expertise on numerical
analysis of hydrodynamic equations,  I. Gornyi for useful discussions and
collaboration on a related project \cite{Gornyi}, and L. Glazman for useful
discussions. While preparing this work for publication, we learnt about a
related activity on the free-fermion problem \cite{Glazman-unpub}. Financial
support by Alexander von Humboldt Foundation (IVP), Israeli Science Foundation
grant 819/10 (DBG), German Israeli Foundation, and DFG priority programs SPP1243 and SPP1285 is gratefully acknowledged. 

\appendix

\section{Hydrodynamics defined by Eqs.~(\ref{Eq:NoHilbert1}),
(\ref{Eq:NoHilbert2}):  Solitons and periodic solutions }
\label{app_Haldane}

In this Appendix we analyze properties of the classical hydrodynamics defined by
Eqs.~(\ref{Eq:NoHilbert1}), (\ref{Eq:NoHilbert2}). Such a theory arises if we
take into account dispersive terms generated by
phenomenological Haldane's formalism, see Sec.~\ref{s3.1}, but neglect the
quantum loop correction. 

As explained in the main text, this turns out to be {\it not} a correct
description of free fermions (since the loop corrections generate
parametrically more important dispersive terms). Nevertheless, the analysis of
this theory is quite illuminating, and we present it in this Appendix.
Furthermore such a theory arises in a fully controllable way in a model of
finite-range interaction with a sufficiently large interaction radius, see
Sec.~\ref{s4.1}.

Let us  focus on  a traveling wave excitations  
\begin{eqnarray}
\label{travel_wave}
\rho(x,t)=\rho(x-Vt)\,,\\
v(x,t)=v(x-Vt)\,.
\end{eqnarray}
Substituting Eq.~(\ref{travel_wave}) into  Eq.~(\ref{Eq:NoHilbert1}),  one finds
\begin{equation}
\label{cont_eq}
-V\partial_x\rho+\partial_x(\rho v)=0\,,
\end{equation}
\begin{equation}
\label{Euler_eq}
 \partial_x\bigg(-Vv+\frac{1}{2}v^2+w \bigg)=0\,.
\end{equation}
We now analyze some simple excitations described  by
Eqs.~(\ref{cont_eq}), (\ref{Euler_eq}).

\subsection{Single soliton}

We start with a solitonic wave. In this case the excitation of the density and
velocity fields are confined  
to a finite region in space, and  the continuity equation (\ref{cont_eq}) yields
\begin{equation}
\label{cont_soliton}
v=V\frac{\rho-\rho_\infty}{\rho}\,.
\end{equation}
Substituting Eq.~(\ref{cont_soliton}) into  Euler equation (\ref{Euler_eq}), one
obtains
\begin{equation}
\label{Euler1}
\rho_x\bigg(\frac{V^2}{2}\frac{\rho_\infty^2-\rho^2}{\rho^2}+\frac{\pi^2\rho^2}{2}-\frac{\pi^2\rho_\infty^2}{2}\bigg)=
\frac{1}{8}\partial_x\left(\frac{\rho_x^2}{\rho}\right)\,. 
\end{equation}
%
Since both sides of the equation are full derivative with respect to $x$, the order of this equation can be easily 
reduced, yielding 
\begin{equation}
\rho_x^2=4\bigg[-V^2\rho_\infty^2-V^2\rho^2+\frac{\pi^2}{3}\rho^4-\pi^2\rho_\infty^2\rho^2+4E\rho\bigg]\,.
\end{equation}
Here  $E=\rho_\infty V^2/2+\pi^2\rho_\infty^3/6$ is a  constant of integration. 
Defining $\xi=\rho-\rho_\infty$, one obtains
\begin{equation}
\int\frac{d\xi}{\xi \sqrt{a+b\xi+c\xi^2}}=2x\,,
\end{equation}
where  $a=\pi^2\rho_\infty^2-V^2$, $b=4\pi^2\rho_\infty/3$, $c=\pi^2/3$. Performing the
integral over $\xi$, we  find the solitonic solution
\begin{eqnarray}
\xi(x)=\frac{4az}{(z-b)^2-4ac}\,,\\
z=\frac{2a+b\xi_0}{\xi_0}e^{-2\sqrt{a}|x|}\,,
\end{eqnarray}
where $\xi_0<0$ is the largest (smallest by absolute value) root of $a+b\xi+c \xi^2$. 
We note that  soliton propagate with the velocity smaller than the velocity of
sound ($V<\pi\rho_\infty$), 
i.e.  is a hole-like excitation from the fermionic point of view.
The charge of a soliton 
\begin{equation}
 q=\frac{\sqrt{3}}{2\pi}\log\frac{2-\sqrt{3(1-\widetilde{V}^2)}}{2+\sqrt{3(1-\widetilde{V}^2)}}\,, \qquad 
\widetilde{V}=\frac{V}{\pi \rho_\infty}
\end{equation}
is  less than unity and is not quantized. 

\subsection{Periodic wave}
 
Periodic solutions of Eqs.~(\ref{cont_eq}), (\ref{Euler_eq}) can be conveniently
parametrized by  $\lambda_4>\lambda_3 >\lambda_2 >\lambda_1$:
\begin{equation}
(\partial_x\rho)^2=\frac{4\pi^2}{3}(\rho-\lambda_1)(\rho-\lambda_2)(\rho-\lambda_3)(\rho-\lambda_4)\,,
\end{equation}
where  parameters $\lambda$  satisfy the constraint $\lambda_1+\lambda_2+\lambda_3+\lambda_4=0$,
and are related to the velocity of the wave according to
\begin{equation}
V^2=\frac{\pi^2}{3}\left(\lambda_2^2+\lambda_3^2+\lambda_4^2+\lambda_2\lambda_3+\lambda_2\lambda_4+\lambda_3\lambda_4 \right).
\end{equation}
The Euler equation, Eq. (\ref{Euler_eq}),  can thus be rewritten as
\begin{equation}
\frac{d\rho}{\sqrt{(\lambda_4-\rho)(\lambda_3-\rho)(\rho-\lambda_2)(\rho-\lambda_1)}}
=\sqrt{\frac{4\pi^2}{3}}dx\,.
\end{equation}
Integrating this equation, we find
\begin{eqnarray}
&& \rho(x)=\frac{\lambda_1(\lambda_4-\lambda_2)[1-{\rm dn}^2(y;k)]
-\lambda_2(\lambda_4-\lambda_1)}{(\lambda_4-\lambda_2)[1-{\rm dn}^2(y;k)]
-(\lambda_4-\lambda_1)}\,. \nonumber\\ &&
\end{eqnarray}
Here ${\rm dn}(y;k)$ is Jacobi elliptic function,
$\nolinebreak{y=x\sqrt{(\lambda_4-\lambda_2)(\lambda_3-\lambda_1)\pi^2/3}}$,
and the elliptic modulus $k$ is given by
\begin{equation}
k^2=\frac{(\lambda_3-\lambda_2)(\lambda_4-\lambda_1)}{
(\lambda_3-\lambda_1)(\lambda_4-\lambda_2)}\,.
\end{equation}

The density oscillates within the interval $\lambda_2 \le
\rho  \le \lambda_3$.
The limit $\lambda_1<\lambda_2< \lambda_3  \rightarrow \lambda_4$  corresponds
to trains of well separated dips (which are nothing but solitons considered
above), where the size of each dip $d$ is much shorter then the distance between
the neighboring dips $L$. In this regime the width of the dip can  be related to
the density amplitude $\delta \rho$  as   
\begin{equation}
d \sim \frac{1}{\sqrt{\delta \rho\, \rho_\infty}} \,.
\end{equation}
The number of particles carried by a dip can be estimated as 
\begin{equation}
q =d\delta \rho  \sim \sqrt{\delta \rho/\rho_\infty} \leq 1\,.
\label{soliton-charge-KDV}
\end{equation}

The limit  $\lambda_1< \lambda_2 \rightarrow \lambda_3 <\lambda_4$ describes a
small-amplitude periodic wave. 
The amplitude of the wave is s $\lambda_3-\lambda_2$, and its wave
length is
$L=\sqrt{3/4\pi^2(\lambda_4-\lambda_3)(\lambda_2-\lambda_1)}$. 
Thus the number of electrons carried  by each ``ripple'' (period) of the wave
is 
\begin{eqnarray}
q \sim \frac{\lambda_3-\lambda_2}{\sqrt{(\lambda_4-\lambda_3)(\lambda_2-\lambda_1)}}\ll \sqrt{\frac{\delta \rho}{\rho_\infty}} \ll 1\,.
\end{eqnarray}

It is worth mentioning that the theory considered in the present Appendix
bears a close connection with the KdV equation of the classical hydrodynamics.
This is because the regularizing term in the hydrodynamic equations has a
similar (third-derivative) structure in both cases.

\section{Modulation equations for hydrodynamics defined by
Eqs.~(\ref{Eq:NoHilbert1}), (\ref{Eq:EntalphyFinal})}
\label{Ap:ModulationEq}

In this section we address the issue of modulation equations for the
hydrodynamic system (\ref{Eq:NoHilbert1}), (\ref{Eq:EntalphyFinal}). Within the
framework of Whitman modulation theory,
one promotes the single-phase (periodic) wave (\ref{Eq:SinglePhaseWave}) to an
Anzats
\begin{equation}
 \phi=\widetilde{\theta}(x, t)+\Phi(\theta(x, t), x, t)
\end{equation}
 and identifies the parameters of the single-phase wave with the derivatives of
the phases $\theta$ and $\widetilde{\theta}$: 
\begin{eqnarray}
\rho_0=\partial_x \widetilde{\theta}\,, \qquad
\gamma=-\partial_t\widetilde{\theta} \,,
\label{Eq:rho_gamma_def}
\\
k=\partial_x \theta\,, \qquad \omega=-\partial_t\theta \,.
\label{Eq:k_omega_def}
\end{eqnarray}
Obviously, parameters defined in this way satisfy the continuity equations
\begin{equation}
 \partial_t\rho_0+\partial_x\gamma=0\,, \qquad \partial_t k+\partial_x \omega=0
\,.
\label{Eq:Continuty}
\end{equation}

\begin{figure}
 \includegraphics[width=230pt]{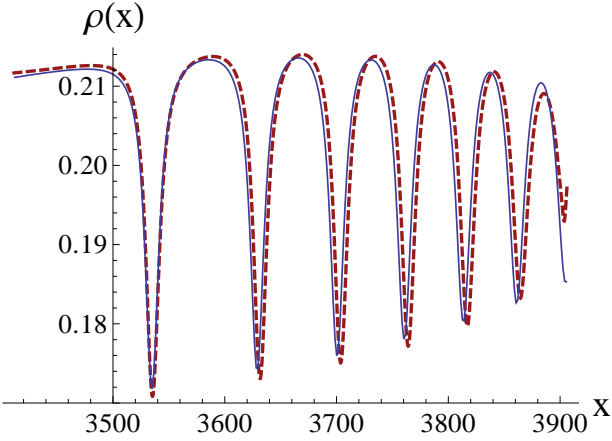}
\caption{\small Comparison of predictions of the modulation theory (dashed line)
and numerical simulations (full line) for the hydrodynamic theory
(\ref{Eq:NoHilbert1}), (\ref{Eq:EntalphyFinal}). A perfect agreement between the
analytic and numerical results is observed. 
  }
\label{Fig:ModulationTheory}
\end{figure}

To derive the modulation equations, we substitute the single-phase wave
(\ref{Eq:SinglePhaseWave}), (\ref{Eq:Polychronakos1}), 
(\ref{Eq:Polychronakos2}), (\ref{Eq:Polychronakos3}) into the Lagrangian $L_p$,
neglect derivatives of the modulation parameters, 
and average the result over a period of oscillations.
We get
\begin{multline}
 \langle L_p\rangle=
-\frac{1}{6} \pi ^2 \rho_1^3+\frac{\gamma ^2}{2 \rho_1}+\frac{\gamma  \omega }{2 \pi \rho_1}
+\frac{\omega ^2}{8 \pi ^2 \rho_1}+\\\frac{k^3}{48 \pi }-\frac{\omega ^2}{4 \pi  k}+
\sigma \left(-\frac{\rho_1 \omega }{2}+\frac{\gamma  k}{2}+\frac{k\omega }{4 \pi }\right)\,.
\label{Eq:L_average}
\end{multline}
Here, $\sigma=\sign\left(k\gamma-\omega\rho_0\right)$ and $\rho_1$ is given by (\ref{Eq:Polychronakos3}).
We now vary the averaged Lagrangian (\ref{Eq:L_average}) with respect to phases
$\widetilde{\theta}$ and $\theta$, 
keeping in mind the relations (\ref{Eq:rho_gamma_def}), (\ref{Eq:k_omega_def}).
This yields
\begin{eqnarray}
 \partial_x \frac{\partial \langle L_p\rangle }{\partial k}+\partial_t \frac{\partial \langle L_p\rangle }{\partial\omega}=0\,.\\
 \partial_x \frac{\partial \langle L_p\rangle }{\partial \rho_0}+\partial_t \frac{\partial \langle L_p\rangle }{\partial\gamma}=0\,.
\end{eqnarray}
These two equations, together with the continuity equations, constitute four
equations for the four unknown parameters. Writing them explicitly and
performing a change of variables according to
Eqs.~(\ref{Eq:k}), (\ref{Eq:omega}), (\ref{Eq:rho0}), (\ref{Eq:gamma}), we
arrive at Eq.~(\ref{Eq:Riemann}).
 
Figure~\ref{Fig:ModulationTheory} demonstrates the density in the shock region
predicted by modulation theory together with the result of numerical
simulations. We observe a perfect agreement between the analytical and
numerical results.  

\section{Modulation theory and soliton trains in Calogero model}
\label{Ap:ModulationEqS}
In Appendix~\ref{Ap:ModulationEq} we have discussed in detail the modulation theory of hydrodynamic equations 
(\ref{Eq:NoHilbert1}), (\ref{Eq:EntalphyFinal}) generated by the ``particle'' Lagrangian $L_p$, Eq.~(\ref{Eq:JevickiLagrangian}). Let us now briefly address the modulation equations for the theory defined by the 
``hole'' Lagrangian $L_h$ and their solution for the upward density perturbation in the initial state. 
This issue is relevant for the description of the solitonic train emerging from the positive density perturbation 
in the repulsive Calogero fluid (see Sec.~\ref{s4.2}).

The starting point for the modulation theory is a single-phase periodic wave. Its form can be obtained from 
Eqs~(\ref{Eq:Polychronakos2}), (\ref{Eq:Polychronakos3}), (\ref{rho-BO}) via the replacement $k\rightarrow-k$ and 
$\omega\rightarrow-\omega$. 
The modulation equations can now be derived exactly in the same way as in Appendix~\ref{Ap:ModulationEq}. 
The result reads
\begin{eqnarray}
 k=u_3-u_2\,,
\label{Eq:kS}
\\
\omega=\frac{1}{2}\left(u_3^2-u_2^2\right)\,,
\label{Eq:omegaS}\\
\rho_0=\frac{-u_0+u_1-u_2+u_3}{2\pi}\,,
\label{Eq:rho0S}\\
\gamma=\frac{-u_0^2+u_1^2-u_2^2+u_3^2}{4\pi}\,,
\label{Eq:gammaS}
\end{eqnarray}
with the Riemann invariants $u_i$ satisfying
\begin{equation}
 \partial_t u_i+u_i\partial_x u_i=0\,.
\end{equation}

Finally, applying the boundary conditions (\ref{Eq:ModBoundary1}), (\ref{Eq:ModBoundary2}) at the edges of shock region 
one finds the Riemann invariants  in term of the branches of Fermi momentum $p_F^{(i)}$ (see Fig.~\ref{Fig:pFBolzman})
\begin{eqnarray}
 u_i=p_F^{(i)}\,,\qquad i=1,2,3 \,,
\label{Riemann1}
\\
u_0=-p_\infty\,.
\end{eqnarray}
Here we have chosen labeling of Riemann invariants $u_i$ such that it exactly corresponds [see Eq.~(\ref{Riemann1})]
to our notations for Fermi momentum branches, as was also the case for the Lagrangian $L_p$, Eq.~(\ref{Eq:ui}).
Note that the relations between the  parameters $k$, $\omega$, $\rho_0$, $\gamma$ and the Riemann invariants differ in the two cases by cyclic permutation of $u_1$, $u_2$ and $u_3$, cf. Eqs.~(\ref{Eq:k}), (\ref{Eq:omega}), (\ref{Eq:rho0}), (\ref{Eq:gamma}) and Eqs.~(\ref{Eq:kS}), (\ref{Eq:omegaS}), (\ref{Eq:rho0S}), (\ref{Eq:gammaS}).

Contrary to the case of ``particle'' Lagrangian $L_p$ considered in the previous section Eqs.~(\ref{Eq:kS}), (\ref{Eq:omegaS}), (\ref{Eq:rho0S}) and (\ref{Eq:gammaS}) predict that the wave vector $k$ vanishes at the leading edge $x_l$ and the density perturbation decays into Lorentzian-shaped solitons 
\begin{eqnarray}
 \rho(x, t)=\rho_\infty+\frac{1}{\pi}\frac{A}{A^2+(x-Vt)^2}\,,\\
  A=\frac{p_F^{(1)}+p_\infty}{2\left(p_F^{(2)}+p_\infty\right)\left(p_F^{(3)}-p_F^{(1)}\right)}\,,\\
  V=\frac{p_F^{(3)}+p_F^{(2)}}{2}\,.
\end{eqnarray}
In  the limit $\Delta\rho\ll \rho_\infty$ this simplifies to
\begin{equation}
 \rho(x, t)=\rho_\infty+\frac{2\Delta\rho}{1+4\pi^2\Delta\rho^2 (x-V_F t)^2}
\end{equation}


\begin{thebibliography}{99}
\bibitem{haldane-footnote} At present the status of Haldane's theory is not
fully understood. We believe that it is 
a phenomenological method that correctly captures the essential features
of charge discreetness as well as of single-particle spectrum curvature.
It is not clear yet, whether it can be accurately derived even for a free
fermion case starting from the first principles.
Note, that the derivation of quantum Hamiltonian(\ref{Eq:QuntumHamiltonian})  
was  done  using  the  collective variables approach \cite{Sakita}.
Though the derivation appears  solid, it contains a very delicate
point\cite{Gornyi}.
In deriving Eq.~(\ref{Eq:QuntumHamiltonian}),  all  Fourier modes of the density
($\rho_k$) were assumed to be independent, 
$\frac{\partial \rho_q}{\partial \rho_k} =\delta_{k,q}$.
While this obviously  holds for an infinite number of particles, the limit  $N
\to \infty$ is delicate. 
Indeed, for any finite number $N$, only the lowest  $N$ modes of the density can
be considered as independent, 
while all the higher harmonics needs to be expressed in terms of the lower ones.
Thus, to treat this limit carefully\cite{Gornyi} one needs to derive the
Hamiltonian for a large but finite number of particles, 
and only then to take the limit $N \to \infty$.
\bibitem{footnote-spectrum} While the one-loop analysis of $\rho^3$ theory
performed in Refs.~\onlinecite{samokhin98,aristov07} generates the correct scale
for spectral broadening of bosonic excitations, it does not produce sharp
boundaries (\ref{spectrum-branches}) of the spectrum. Within such an approach,
the branches (\ref{spectrum-branches}) should emerge as a result of 
resummation of the whole perturbative expansion.
\bibitem{footnote-cutoff} In the estimation of zero momentum interaction $V_0$ we have used  
natural ultraviolet cutoff $1/\rho_\infty$.
\bibitem{Remark1} We have checked that, as expected, our results do not depend
on the density $\rho_\infty$ (up to renormalization of Fermi velocity).
\end{thebibliography}
\end{document}